\documentclass[]{article}

\usepackage{graphicx}
\usepackage{epstopdf}
\usepackage{ifpdf} %%% pdf package
\usepackage{hyperref}
\usepackage{xcolor}
\hypersetup{
	colorlinks,
	linkcolor={red!90!black},
	citecolor={green!80!black},
	urlcolor={blue!80!black}
}

\usepackage{epstopdf} 
\usepackage{amsmath,amssymb}
\usepackage{mathrsfs}
\usepackage{tabularx} 	
\usepackage{gensymb}	% for degree sign
\usepackage[version=3]{mhchem}
\usepackage{units}

\usepackage{fullpage, natbib}
\newcommand{\footremember}[2]{%
	\footnote{#2}
	\newcounter{#1}
	\setcounter{#1}{\value{footnote}}%
}
\newcommand{\footrecall}[1]{%
	\footnotemark[\value{#1}]%
} 

\usepackage[edtable,longtable]{lineno} %%% number lines
\nolinenumbers

\usepackage{setspace} %%% Spacing
\usepackage{booktabs} %%% for much better looking tables
\usepackage{graphicx}

\usepackage{comment}
\setkeys{Gin}{draft=false} %%%  Uncomment the following command to allow illustrations to print when using Draft:

\usepackage{enumitem} %%% enumeration
\setlist{nolistsep}
\usepackage{mdframed} %%% used here to create the box around the disclaimer
\usepackage{caption}
\usepackage{csquotes}
\usepackage{longtable}
\pdfoutput=1
\begin{document}
	
	%%%
	%%% DEFINITIONS
	%%%
	\newcommand{\fixme}[1]{ { \bf \color{red}FIX ME \color{black} #1 } }
	\newcommand{\fig}{figures/}
	%%% adds space in tables
	\newcommand\T{\rule{0pt}{2.6ex}}         % = `top' strut
	\newcommand\B{\rule[-0.9ex]{0pt}{0pt}}   % = `bottom' strut

	%%%%%%%%%%%%%%%%%%%%%%%%%%%%%%%%%%%%%%%%%%%%%%%%%%%%%%%%%%%%%%%%%
	%%% TITLE WITH AUTHORS

	\title{Estimating Flow Rates through Fracture Networks\\ using Combinatorial  Optimization} 
	\author{
		A. Hob\'e \footremember{add1}{ETH Zurich, Geothermal Energy and Geofluids, Institute of Geophysics, Zurich, Switzerland}\and 
		D. Vogler \footrecall{add1}
		\footremember{add2}{ETH Zurich, Transport Processes and Reactions Laboratory, Institute of Process Engineering, Zurich, Switzerland} \and %\corref{cor1}
		M. P. Seybold \footremember{add3}{University of Stuttgart, Institute of Formal Methods in Computer Science, Stuttgart, Germany}\and 
		A. Ebigbo \footrecall{add1} \and 
		R. R. Settgast\footremember{add4}{Lawrence Livermore National Laboratory, Atmospheric, Earth and Energy Division, Livermore, CA, USA}  \and 
		M. O. Saar\footrecall{add1}
	}
	\date{}
	
	%%%%%%%%%%%%%%%%%%%%%%%%%%%%%%%%%%%%%%%%%%%%%%%%%%%%%%%%%%%%%%%%%

\vspace{1cm}
\maketitle

%%%%%%%%%%%%%%%%%%%%%%%%%%%%%%%%%%%%%%%%%%%%%%%%%%%%%%%%%%%%%%%%%
%%%
%%% ABSTRACT
%%%
\begin{abstract}
	%%%%%%%%%%%%%%%%%%%%%%%%%%%%%%%%%%%%%%%%%%%%%%%%%%%%%%%%%%%%%%%%%
	%
To enable fast uncertainty quantification of fluid flow in a discrete fracture network (DFN), we present two approaches to quickly compute fluid flow in DFNs using combinatorial optimization algorithms. 
Specifically, the presented Hanan Shortest Path Maxflow (HSPM) and Intersection Shortest Path Maxflow (ISPM) methods translate DFN geometries and properties to a graph on which a max flow algorithm computes a combinatorial flow, from which an overall fluid flow rate is estimated using a shortest path decomposition of this flow.
The two approaches are assessed by comparing their predictions with results from explicit numerical simulations of simple test cases as well as stochastic DFN realizations covering a range of fracture densities. 
Both methods have a high accuracy and very low computational cost, which can facilitate much-needed in-depth analyses of the propagation of uncertainty in fracture and fracture-network properties to fluid flow rates.
%}
%
\end{abstract} % 139 words 
% Should be between 100 - 150 words according to author's guide.

%%%%%%%%%%%%%%%%%%%%%%%%%%%%%%%%%%%%%%%%%%%%%%%%%%%%%%%%%%%%%%%%%
%%%
%%% KEYPOINTS
%%%

%%%%%%%%%%%%%%%%%%%%%%%%%%%%%%%%%%%%%%%%%%%%%%%%%%%%%%%%%%%%%%%%%
%\end{frontmatter}

%%%%%%%%%%%%%%%%%%%%%%%%%%%%%%%%%%%%%%%%%%%%%%%%%%%%%%%%%%%%%%%%%
%%%
%%% INTRODUCTION
%%%
\section{Introduction} \label{sec:intro}
%%%%%%%%%%%%%%%%%%%%%%%%%%%%%%%%%%%%%%%%%%%%%%%%%%%%%%%%%%%%%%%%%
%
Fractured rock reservoirs are important for society.
They play a role in geothermal energy applications \citep{rybach2000swiss,tester2006future,saar2011geothermal,bertani2016geothermal,lund2016direct}, unconventional oil and gas recovery \citep{van1982fundamentals,kissinger2013hydraulic}, radioactive waste disposal \citep{Outters2000,Tsang2015}, and drainage of tunnels \citep{wilhelm2003geothermal,or2005seepage}, to name only a few. 
In all these applications, it is vital to have an accurate understanding of the way fluids flow through such reservoirs, where fractures can serve as preferential fluid pathways with flow velocities that can be orders of magnitude greater than in the rock matrix. 
This flow behavior depends on the mechanisms of flow within each individual fracture and on the geometric configuration of the fractures in the network. 
Flow in a single fracture is strongly affected by its aperture distribution, as it governs the transmissivity of the fracture \citep{brown_1987,watanabe_2008,vogler_2016b,vogler_2018}. 
The fracture transmissivity is often represented using the cubic law -- or some variation thereof \citep{ZIMMERMAN1996,liu2016} -- which states that the fracture transmissivity is proportional to the cube of the average aperture of the fracture. 
At the reservoir scale, more connected fractures yield more potential flow paths, and hence a higher effective permeability of the rock \citep{ebigbo2016inclusion}. 
This connectivity of fractures, or planar features in 3D space, is dependent on the dimensions (e.g. aspect ratio), orientation, and size distribution as well as the spacing of these planar features, as described by percolation theory \citep[e.g.][]{saar2002continuum,walsh2008magma,walsh2008numerical,davis2011statistically}. 
Connected fractures form a discrete fracture network (DFN), where the fracture parameters can be distributed constant in space, or uniformly, log-normally, or according to a power law \citep{ZIMMERMAN1996,de_dreuzy_2001_1,de_dreuzy_2001_2,Leung2012}. 
DFNs are often generated stochastically given such parameter distributions \citep{adler2013fractured}. 

However, due to a lack of accessibility, differences in scale, and the complex, heterogeneous, and 3D nature of the problem, it is nearly impossible to obtain sufficient data to adequately characterize a fracture network at depth \citep{clauser1992permeability,ledesert1993geometrical,Faybishenko2000,berkowitz2002characterizing,dowd2009,wilson2015developing}. 
The sensitivity of fluid flow to DFN properties, therefore, needs to be assessed by performing uncertainty quantifications, to aid in decision-making for the mentioned applications.

Currently, most methods for describing flow through fractured reservoirs fall into two categories:
\begin{enumerate}[label=\arabic*)]
\item 
Partial differential equations solvers (referred to here as direct simulations, DS) using the finite element method (FEM), the finite volume method (FVM), or other methods on discrete meshes \citep[e.g.][]{pruess1999tough2,diersch2013feflow,tatomir2013discrete,lang2014permeability,hyman2015dfnworks,settgast2016fully}.
\item
Analytical approximations of the equivalent permeability of the fractured rock, based on fracture and fracture-network properties  \citep[e.g.][]{Oda1985,ZIMMERMAN1996,de_dreuzy_2001_1,de_dreuzy_2001_2,Mourzenko2011,adler2013fractured,Saevik2013,ebigbo2016inclusion,liu2016}. 
\end{enumerate}

DS give an accurate solution for a specific DFN problem, but are computationally expensive \citep{de2013synthetic}.
This expense, combined with the required compute time, is why in-depth quantifications of uncertainties are rarely done employing DS.
The analytical approximations express the equivalent fractured-rock permeability as a function of characteristic fracture network properties such as fracture density, orientation, and aperture, typically within some ranges of validity. 
They thus produce one mean permeability tensor value per set of stochastic fracture-network attributes, giving no information about the uncertainty associated with the obtained values. 
Other information, such as preferential flow paths, are not generated by these methods either. 

In this paper, we present two approaches to rapidly compute fluid flow in DFNs  using existing graph theory algorithms \citep{Edmonds1972,newman2010networks,ahuja1993network}. 
These two approaches enable fast uncertainty quantifications associated with a DFN parameter space.
The Hanan Shortest Path Maxflow (HSPM) and the Intersection Shortest Path Maxflow (ISPM) methods translate DFN geometries and properties to a graph, on which a so-called maximum flow algorithm is used to find a so-called combinatorial flow, from which an overall fluid flow rate is estimated, using a shortest path decomposition of this flow.
The HSPM and ISPM methods mainly differ in their graph extraction methods.
The predictions of overall fluid flow through a DFN, employing these two approaches, are compared with results from direct simulations of simple test cases as well as stochastic DFN realizations, covering a range of fracture densities.
%
%
%
%%%%%%%%%%%%%%%%%%%%%%%%%%%%%%%%%%%%%%%%%%%%%%%%%%%%%%%%%%%%%%%%%
%%%
%%% METHODS
%%%
\section{Methods} \label{sec:methods}
%%%%%%%%%%%%%%%%%%%%%%%%%%%%%%%%%%%%%%%%%%%%%%%%%%%%%%%%%%%%%%%%%
%

%%%%%%%%%%%%%%%%%%%%%%%%%%%%%%%%%%%%%%%%%%%%%%%%%%%%%%%%%%%%%%%%%
%
%%%%%%%%%%%%%%%%%%%%%%%%%%%%%%%%%%%%%%%%%%%%%%%%%%%%%%%%%%%%%%%%%
%%%
%%% SUBSECTION - GRAPH THEORY BACKGROUND
%%%
\subsection{Graph theory} \label{sec:graph_theory} 
%%%%%%%%%%%%%%%%%%%%%%%%%%%%%%%%%%%%%%%%%%%%%%%%%%%%%%%%%%%%%%%%%
Methods that build upon an underlying discrete structure, such as a
graph network, have become very popular in many fields of science in
recent years \citep[e.g.][]{newman2010networks,wittkowski2013single,ahuja1993network,funke2014generalized,seybold2017robust}. 
Since the advent of the Internet, and particularly social networks, hundreds of mathematical tools have been created to investigate network-like structures and phenomena \citep{newman2010networks}.
Any phenomenon that can be described as a network with connection points (nodes or vertices) and relationships between them (edges) can be investigated using these mathematical tools. 
Recent applications of such tools in the geosciences were reviewed by \cite{phillips2015graph}, while \cite{Heckmann2015} specifically reviewed their usage in geomorphology.
Recent examples of graph theory usage in fracture flow problems are as follows: 
\cite{ghaffari2011fluid} used graph theory to describe fracture clusters and fracture zones in 2D fractured media and their influence on permeability. 
\cite{aldrich2017analysis} used graph theory to analyze and visualize the particle-tracking results of flow and transport DS. 
\cite{hyman2017predictions} simplified DS investigations of first passage times of passive tracer particles by reducing the DFN to the most important fractures using graph theory. 
\cite{karra2017modeling} used a graph representation to model flow and transport in DFNs in order to investigate breakthrough times and statistical data of tracer particles. 
The work of \cite{karra2017modeling} will be discussed further in Section~\ref{sec:flowToGraph},  due to its similarities to our study. 
\subsubsection{The maximum flow problem} \label{sec:maxflow}
%\emph{
A \emph{fundamental difference exists between fluid flow, $Q$, and combinatorial flow, $f$,} which will be summarized in Section~\ref{sec:backTranslation}.
All references to flow in Sections~\ref{sec:maxflow} through \ref{other_alg} refer to combinatorial flow on a graph, except when it is specifically stated that fluid flow in fracture networks is meant.
%}
\\

The most important mathematical tool in this study is the Edmonds-Karp algorithm \citep{Edmonds1972} which finds the maximum amount of flow between two nodes of a graph and thus solves the so-called \textit{max flow optimization problem}. See \cite{ahuja1993network} for a discussion 
%proof 
of this algorithm's accuracy and efficiency.

\begin{figure}[h] \centering
\includegraphics[width=0.6\textwidth]{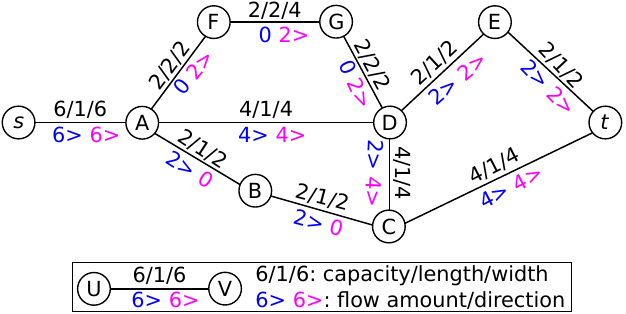}
\caption{A graph representation of a network with a source, $s$, and a sink, $t$. 
Edges are labeled with capacity/length/width values (black) and two valid max flow assignments (blue and magenta). 
Such a flow assignment shows how much flow occurs on the individual edges.
The edges along A,F,G,D for the blue flow assignment and along A,B,C for the magenta flow assignment do not contribute to the final flow result.
Edge directions are omitted here, since both directions bear identical labels.}
\label{exampleGraph}
\end{figure}

To introduce the fundamental definitions and assumptions of the max flow problem, we start with a directed graph network, $G$, comprised of a set of vertices, $V$, and a set of directed edges, $E$, which connect the vertices (a directed edge only has flow in one direction).
The amount of flow, $f : E \to \mathbb{R}^+$, an edge, $e \in E$, can accommodate is fixed in a graph representation. 
This is called the edge capacity, $c : E \to \mathbb{R}^+$, which can be any assigned value and should therefore not be confused with capacitance in electrical circuits.
Given a directed graph, $G = (V,E)$, with non-negative edge capacities, $c : E \to \mathbb{R}^+$, and two vertices, $s,t \in V$ (the chosen source and sink vertices), a mapping, $f : E \to \mathbb{R}^+$, is called a valid $s$-$t$ flow, if, for all edges $e \in E$, the inequality $f(e) \leq c(e)$, and, for all vertices $v \in V \setminus \{s,t\}$, the flow conservation constraint,
\begin{linenomath*}
\begin{align}
\sum_{(\cdot,v) \in E} f(\cdot,v) - \sum_{(v,\cdot) \in E} f(v,\cdot) =0 ,
\end{align}
\end{linenomath*}
holds.
We call $\sum_{(\cdot,t)\in E} f(\cdot,t)$ the \emph{value} of a flow and a flow, $f : E \to \mathbb{R}^+$, with maximum value is called a maximum $s$-$t$ flow.
The maximum flow of both flow assignments in Figure \ref{exampleGraph} (blue and magenta) is, therefore, $\unit[6]{units}$.
These two flow assignments are illustrated to demonstrate how the distribution of flow can differ, while obtaining the same flow-value result.
Several other valid flow assignments exist for this graph.

%%%%%%%%%%%%%%%%%%%%%%%%%%%%%%%%%%%%%%%%%%%%%%%%%%%%%%%%%%%%%%%%
%%% FIGURE GRAPH EXAMPLE
%%%
%%%%%%%%%%%%%%%%%%%%%%%%%%%%%%%%%%%%%%%%%%%%%%%%%%%%%%%%%%%%%%%%%
%
%
%
%
%%%%%%%%%%%%%%%%%%%%%%%%%%%%%%%%%%%%%%%%%%%%%%%%%%%%%%%%%%%%%%%%%
\subsubsection{Edmonds-Karp algorithm} \label{maxflow_algorithm}
\begin{figure}[htbp] 
\centering
a)\includegraphics[width=0.48\textwidth]{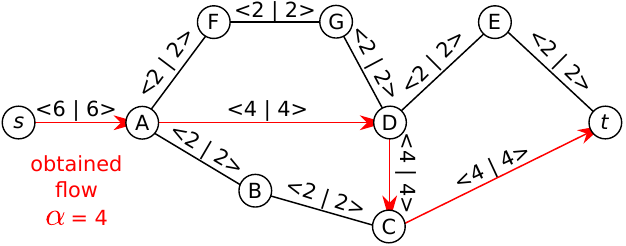}\\
b)\includegraphics[width=0.48\textwidth]{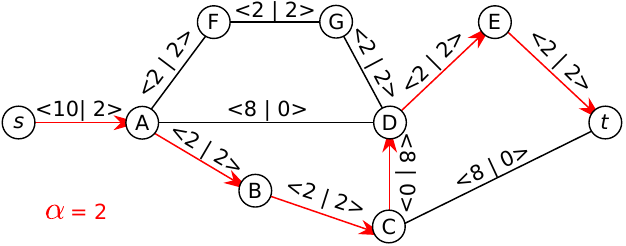}\\
c)\includegraphics[width=0.48\textwidth]{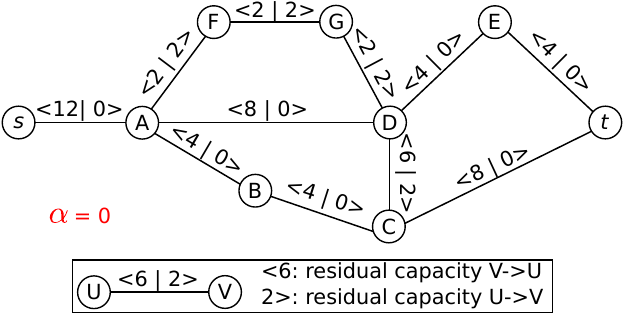}
\caption{
Description of how the blue flow assignment in Figure~\ref{exampleGraph} is found using the corresponding residual graphs with residual capacities assigned to both edge directions.
\textbf{a)} The first augmenting path is found as a shortest path, where all edges have a residual capacity larger than zero.
The smallest residual edge capacity, $\alpha$, encountered on this path is $\unit[4]{units}$.
\textbf{b)} The $\alpha$-value found in (a) has been removed from the edges of the path in (a) and has been added to the edges in the reverse direction so that a subsequent path has the possibility to cancel out this previously obtained flow.
The path shown in this figure does this for the vertical edge $(C,D)$, where it cancels out $\unit[2]{units}$ of flow.
\textbf{c)} The $\unit[2]{units}$ of flow found by the second augmenting path have been removed from the graph in (b).
There are no more possible paths between $s$ and $t$ that have a non-zero capacity. 
}
\label{3graphs}
\end{figure}

%%%%%%%%%%%%%%%%%%%%%%%%%%%%%%%%%%%%%%%%%%%%%%%%%%%%%%%%%%%%%%%%%
%
The Edmonds-Karp algorithm \citep{Edmonds1972} is used in general to find the maximum flow that a given graph network, $G$, can accommodate between a source node, $s$, and a target node, $t$, which can be any two nodes in $G$. 
The Python graph-tool \citep{peixoto-graph-tool-2014}, which we employ for graph representations, provides a version of this algorithm which is called the ``labeling'' algorithm in  \cite{ahuja1993network}.
This uses the ``hop distance'' and two copies of the graph network under investigation. 
The hop distance from the source node corresponds to the minimal number of edges, $e$, that need to be traversed to go from the source node to the node in question.
The initial graph, $G$, is used to store the flow values, $f_e$, found for the individual edges as the algorithm finds new flow paths.
A second graph, $G_r$, (the ``residual'' graph) is used to determine how much flow is still possible on each edge in $G$, referred to as the ``residual edge capacity'', $cr_e = c_e - f_e$.
The residual capacity, $cr_{e'}$, in the opposite direction (indicated by $e'$) of the current flow value is assigned to the edges on $G_r$. This allows the algorithm to cancel part or all of the previously found flow on an edge in $G$. This ensures that the algorithm does not terminate before obtaining the correct flow value.
After the first path in Figure~\ref{3graphs}a is found, for example, the first edge from the source $(s,A)$ in Figure~\ref{3graphs}b has a residual capacity of \unit[2]{units} (i.e.\ $6-4 = 2$) and the reverse edge
$(A,s)$ in Figure~\ref{3graphs}b has been given $4$ additional units.
If a source/target path with positive residual capacities exists, this path can accommodate additional flow units and is, therefore, called an ``augmenting'' path.

%%%%%%%%%%%%%%%%%%%%%%%%%%%%%%%%%%%%%%%%%%%%%%%%%%%%%%%%%%%%%%%%
%%% FIGURE RESIDUAL GRAPH EXPLANATION
%%%
%%%%%%%%%%%%%%%%%%%%%%%%%%%%%%%%%%%%%%%%%%%%%%%%%%%%%%%%%%%%%%%%
%

To find an augmenting path on the residual graph, which will have the shortest hop distance of all possible augmenting paths, the Edmonds-Karp algorithm uses a ``breadth-first'' search, which works as follows:
\setlist[enumerate]{label=\roman*), ref=\theenumii.\roman*}
\begin{enumerate}
\item Starting at the source node, each outward-going edge with positive residual capacity is traversed.
\item The new vertices are then labeled according to the hop distance from the source.
\item From these vertices, the outward-going edges are again traversed, but only to unlabeled vertices.
Any previously labeled vertex would have an equal or shorter hop distance than the current vertex.
\item This continues until the target node is found, or until no more unlabeled nodes can be moved to from the current nodes.
In the latter case, no additional path between the source and target node exists.
\item The path is found by following the labels in descending order from the target node to the source node.
\end{enumerate}
\medskip

The amount of flow, $\alpha$, each augmenting path adds to $G$, corresponds to the smallest $cr_e$ encountered on this path on $G_r$.
This $\alpha$ is then added to the previously obtained values of $f_e$ in $G$ for each edge along the augmenting path, and the residual graph is updated using the new flow.
This process is repeated until no more paths are found on the residual graph, with the    encountered vertices getting new labels during every breadth-first search.
The graph's maximum flow is obtained by summing $f_e$ on the edges connecting to the target node.

%%%%%%%%%%%%%%%%%%%%%%%%%%%%%%%%%%%%%%%%%%%%%%%%%%%%%%%%%%%%%%%%
%%%
%%% TABLE AUGMENTING PATH RESULTS
%%%
\begin{table}[htbp]
	\centering
	\caption{Flow values (initially zero) for individual edges in Figure~\ref{3graphs} for augmenting paths p1 and p2 and final flow results.
	}
	\label{tab_augPath}
    \begin{center}
		\begingroup 
		\setlength{\tabcolsep}{13pt} % Default value: 6pt
\begin{tabular}{llll}
\hline
Edge & p 1   & p 2   & flow \\
\hline
$s$A   & (4,0) & (6,0) & (6,0)\\
AB   & 		 & (2,0) & (2,0)\\
AD   & (4,0) & 		 & (4,0)\\
AF	 & 		 &		 & (0,0)\\
BC   & 		 & (2,0) & (2,0)\\
CD   & (0,4) & (2,4) & (0,2)\\
C\hspace{1pt}$t$   & (4,0) & 		 & (4,0)\\
DE   & 		 & (2,0) & (2,0)\\
DG   & 		 & 		 & (0,0)\\
E\hspace{1pt}$t$   & 		 & (2,0) & (2,0)\\
FG   & 		 & 		 & (0,0)\\
\hline
\end{tabular}
\endgroup
\end{center}
\end{table}
%%%%%%%%%%%%%%%%%%%%%%%%%%%%%%%%%%%%%%%%%%%%%%%%%%%%%%%%%%%%%%%%
%

Table~\ref{tab_augPath} shows the flow on the individual edges in Figure~\ref{3graphs} for the two augmenting paths and the final flow results on those edges. 
The vertical edge CD has flow in two directions as the second augmenting path is found.
In the final result, this is simplified to a positive value in one direction and zero in the other.

The use of augmenting paths, that correspond to successive shortest paths on the residual graph, makes the resulting flow decomposition beneficial for this project.
Although the paths that make up the flow decomposition are not equal to the augmenting paths used to find this flow decomposition, the former tend to inherit this property (shortness) from the latter.
The shortest path decomposition of the final flow (c.f. Section~\ref{sec:backTranslation}) finds the maximum amount of fluid flow, $Q$, for each successive path, because the resistance to fluid flow increases with path length (Eq~\ref{eq:cubiclaw}). 
The off-the-shelf version of the Edmonds-Karp algorithm makes use of the hop distance to find the shortest path, instead of the Euclidean length.
This is unfortunate, because the path with the shortest hop distance does not correspond to the path with the shortest Euclidean length in all cases (there is no difference for the chosen paths in Figure~\ref{3graphs}b).
Its use here, as part of the existing Edmonds-Karp algorithm, simplifies this proof-of-concept investigation, compared to implementing a new algorithm.
Although the maximum flow value of a graph is unique, the distribution of the possible paths is not, as seen when comparing the blue values and the magenta values in Figure~\ref{exampleGraph}.
The way the Edmonds-Karp algorithm distributes the flow over possible paths is influenced by the order in which nodes and edges are defined on the graph, which can also be accomplished by switching the source and target nodes.
\subsubsection{Other max flow algorithms} \label{other_alg}
Many other algorithms exist that solve the max-flow problem \citep{goldberg1989network,boykov2004experimental,ahuja1993network}, of which the Python graph-tool \citep{peixoto-graph-tool-2014} provides two (next to Edmonds-Karp). 
We tested the ``push-relabel" algorithm and the ``Boykov-Kolmogorov" algorithm, but did not deem them suitable for the purposes of this work, as the combinatorial flow assignments tend to use paths that are non-physical for the fluid flow case.
%
%
%
%

%%%%%%%%%%%%%%%%%%%%%%%%%%%%%%%%%%%%%%%%%%%%%%%%%%%%%%%%%%%%%%%%%
%%%
%%% APPLICATION OF GRAPH THEORY TO FRACTURE NETWORKS
%%%
\subsection{Application of graph theory to fracture networks} \label{appl}
%%%%%%%%%%%%%%%%%%%%%%%%%%%%%%%%%%%%%%%%%%%%%%%%%%%%%%%%%%%%%%%%%

Several fundamental questions are associated with applying the available mathematical tools to solve for fluid flow in DFNs. Firstly, the geometries and properties of a DFN need to be translated to a graph representation, consisting of nodes and edges (discussed in Section~\ref{sec:flowToGraph}).
The Edmonds-Karp algorithm is applied on the produced graph to obtain a distribution of combinatorial flow (as seen in Figure~\ref{exampleGraph}).
This combinatorial flow then needs to be translated back to a fluid flow result (discussed in Section~\ref{sec:backTranslation}).
The following assumptions were used for these steps and for the subsequent result assessment employing DS.

Fluid flow is restricted to the fractures in each DFN and matrix flow is neglected, assuming that the matrix permeability is very low. 
This assumption is usually valid for crystalline rock and shale \citep[e.g.][]{Outters2000,Leung2012}.
Although fractures can have complex geometries and fracture surfaces are heterogeneous in nature \citep{brown_1987,watanabe_2008,vogler_2016b,vogler_2017b,vogler_2018}, this work defines fractures as 2D planes with a length, $l$, a width, $W$, and an aperture, $a$, for simplicity. 
Furthermore, for the purposes of this study, these fracture planes are all parallel to either the $x-y$, $x-z$, or $y-z$ plane of the principal coordinate system. 
Such orthogonal fractures are not uncommon in geological systems \citep{bock1980fundamentale,dunne1990orthogonal,bai2002orthogonal}.
These modeling choices do not reflect a limitation of any method employed in this study, but merely simplify the fracture-network generation and the meshing process, required for the DS verification runs. 
The parameters used for simple and stochastically generated DFNs are shown in Table \ref{tab_sim_param}.
%
%%%%%%%%%%%%%%%%%%%%%%%%%%%%%%%%%%%%%%%%%%%%%%%%%%%%%%%%%%%%%%%%%
%%%
%%% TABLE - SIMULATION PARAMETERS
%%%
\begin{table}[htbp]
	\centering
	\caption{Parameters used during fracture network generation.}
	\label{tab_sim_param}
	\begin{center}
		\begingroup
			\setlength{\tabcolsep}{20pt} % Default value: 6pt
				\begin{tabular}{ c  c   c c c c }  
				\hline
				$\rho$  &  $\mu$  &  $g$  &  $a$  &  $\Delta P$  &  $dx,\,dy,\,dz$ \\[2pt] 
				$\unit{kg/m^3}$ & $\unit{Pa \cdot s}$ & $\unit{m/s^2}$ & $\unit{m}$ & $\unit{MPa}$ & $\unit{m}$ \\[2pt]
				\hline
				1000 & 0.001 & 9.81 & $10^{-5}$ & 1.0 & 0.2 \T \\[2pt]
				\hline
				\end{tabular}
		\endgroup
	\end{center}
\end{table}
%%%%%%%%%%%%%%%%%%%%%%%%%%%%%%%%%%%%%%%%%%%%%%%%%%%%%%%%%%%%%%%%%

These simplifications allow the usage of the cubic law for saturated fracture flow \citep{ZIMMERMAN1996} for fluid flow computations:
\begin{linenomath*}
\begin{equation}
Q = K \cdot (\Delta P - \rho g \Delta z) \hspace{1cm} \textrm{with} \hspace{1cm} K = \frac{w a^3}{12 \mu L},
\label{eq:cubiclaw}
\end{equation}
\end{linenomath*}
where, in this study, $L$ and $w$ represent the length and width attributed to the given edge and $\Delta z$ is the vertical distance (against gravity) traversed by the fluid. 
The definition of the hydraulic conductance, $K$, in EQ.~\ref{eq:cubiclaw} is the same in this study, as that of \cite{ZIMMERMAN1996}.  
$K$ has units of $m^3/(Pa\cdot s)$, denotes the volumetric flow rate obtained for a given pressure difference, and
is very similar to the inverse of resistance in Ohm's law for electrical currents.

%%%%%%%%%%%%%%%%%%%%%%%%%%%%%%%%%%%%%%%%%%%%%%%%%%%%%%%%%%%%%%%%%
%%%
%%% TRANSLATING FROM FLUID FLOW TO A GRAPH REPRESENTATION
%%% 
\subsubsection{Translating from fluid flow to a graph representation} \label{sec:flowToGraph}
%%%%%%%%%%%%%%%%%%%%%%%%%%%%%%%%%%%%%%%%%%%%%%%%%%%%%%%%%%%%%%%%%
To translate from fluid flow in a DFN to a graph representation, entities need to be chosen to be represented as nodes and corresponding flow values between such entities need to be defined to be represented as edges.
This paper describes two possible entities as candidates to be represented as nodes:
The Hanan Shortest Path Maxflow (HSPM) method uses fracture segments. The Intersection Shortest Path Maxflow (ISPM) method uses fracture intersections.
Both methods use the cubic law (Eq.~\ref{eq:cubiclaw}) to assign weights to the edges, but do so differently.
A step-by-step explanation of both methods is given in the following sections.

\begin{figure}[!ht]
\centering
a) \includegraphics[width=0.5\textwidth]{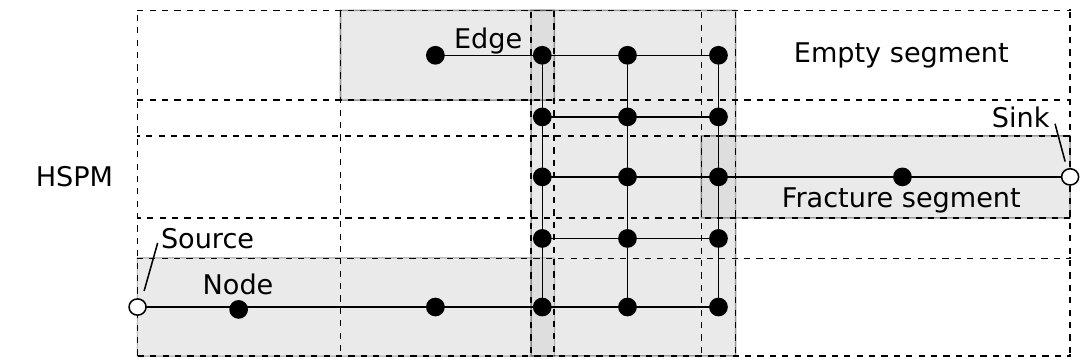} \\
b) \includegraphics[width=0.5\textwidth]{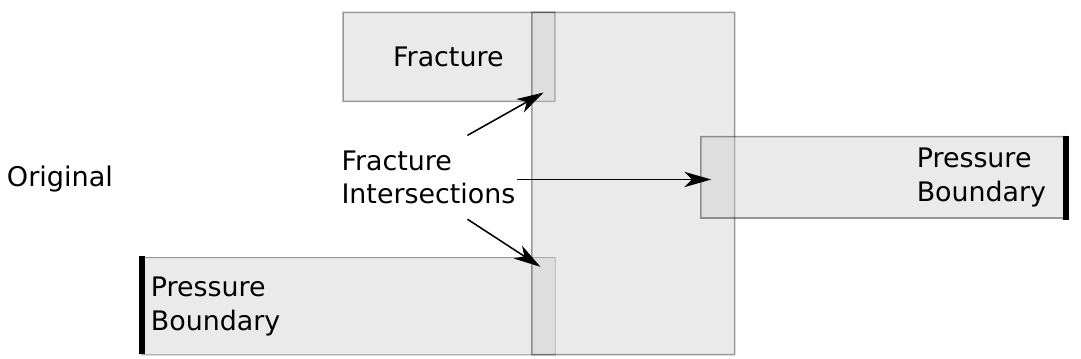}\\
c) \includegraphics[width=0.5\textwidth]{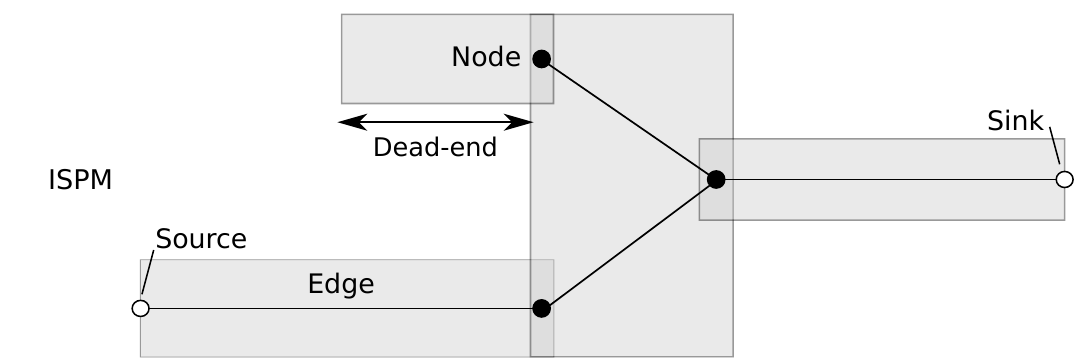} 
\caption{
Graph extraction results of \textbf{a)} the HSPM and \textbf{c)} the ISPM  method for the original DFN in \textbf{(b)}.
Each graph consists of nodes and edges connecting these nodes.
Source and sink nodes correspond to the pressure boundaries.
}
\label{segmentation}
\end{figure}

The HSPM method (Figure~\ref{segmentation}a) 
uses the fracture coordinates to slice the entire domain into unevenly shaped boxes (segments). 
The segments that lie on fractures will be assigned a vertex with the coordinate of the vertex being the center of the box. 
Neighboring vertices are connected to each other by edges. The result is called a Hanan grid \citep{Zachariasen00acatalog}. 
The source and the sink of the graph correspond to the pressure boundaries of the DFN. 
When multiple fractures intersect with the pressure boundary, they are connected to a so-called \textit{super-source} or \textit{super-sink}. 
The edges, connecting to the super-source and super-sink, are given a large capacity value to simulate instantaneous flow.
In 3D, additional segmentation of this example could happen due to fractures above or below these fractures.
Using Eq.~\eqref{eq:cubiclaw}, the capacity value, $c_e$, assigned to each edge, corresponds to the hydraulic conductance, $K_e$, of a fracture with aperture, $a_e$, width, $w_e$, and length, $L_e$.
HSPM uses the width of the segment perpendicular to the edge, the distance between the two centroid points, and constant aperture values, though varying apertures could be implemented.
Edges that would have a width equal to or less than the fracture aperture are ignored to speed up this method.
%
%

%%%%%%%%%%%%%%%%%%%%%%%%%%%%%%%%%%%%%%%%%%%%%%%%%%%%%%%%%%%%%%%%%
%%%
%%% FIGURE SEGMENTATION DESCRIPTION
%%%
%%%%%%%%%%%%%%%%%%%%%%%%%%%%%%%%%%%%%%%%%%%%%%%%%%%%%%%%%%%%%%%%%

The ISPM method uses fracture intersections as nodes and connects these directly along the longest dimension of each fracture (Figure \ref{segmentation}c). 
Dead-end fractures that do not lie between two intersections are ignored, as they do not contribute to the flow path \cite[e.g][]{Leung2012}. 
As in the HSPM case, this method assigns a (super-) source and (super-) sink to the fractures intersecting the pressure boundary. 
A large capacity value is employed for instantaneous flow simulation in two different cases in the ISPM method. Firstly, it is implemented for the edges connecting to a super-source/sink and secondly, it is implemented, where multiple fractures intersect at the same place, because this creates nodes with the same coordinates for each intersecting fracture pair and connecting edges with a length of zero.

For the ISPM method, the width of the fracture (the second-largest dimension of the fracture), $W$, is used for $w$ in Eq.~\eqref{eq:cubiclaw}. 
This makes sense for fractures with pronounced eccentricity.
For DFNs with such fractures, there is a greater likelihood of fluid flowing along the length of the fractures between intersections. 

There are some similarities and some differences between the ISPM method presented here and that of \cite{karra2017modeling}, which warrant a brief summary.
\cite{karra2017modeling} also represent fracture intersections as nodes in the graph extraction, but all fracture intersections, lying on the same fracture, are connected with each other and the width of the edge is chosen as the arithmetic mean of the sizes of two connected intersections.
They use the graph Laplacian \citep{newman2010networks} of the obtained graph to solve for fluid pressure on the vertices. 
This information is used to calculate fluid velocities on individual edges, which is subsequently used in a particle-tracking algorithm to solve for transport.
The determined breakthrough times of passive tracer particles were up to an order of magnitude less, compared to their DS results, which the authors attribute to an underestimation of the pressure gradients across intersections by their algorithm.

%%%%%%%%%%%%%%%%%%%%%%%%%%%%%%%%%%%%%%%%%%%%%%%%%%%%%%%%%%%%%%%%%
%%%
%%% TRANSLATING THE RESULT FROM THE GRAPH REPRESENTATION BACK TO FLUID FLOW
%%%
\subsubsection{Translating the results from the graph representation back to fluid flow} \label{sec:backTranslation}
%%%%%%%%%%%%%%%%%%%%%%%%%%%%%%%%%%%%%%%%%%%%%%%%%%%%%%%%%%%%%%%%%
After applying the Edmonds-Karp algorithm (Section~\ref{maxflow_algorithm}), the Python graph-tool does not provide possible fluid flow paths directly.
It only provides the flow results for the edges of the graph as part of the calculated flow assignment (e.g.\ the blue flow in Figure~\ref{exampleGraph}).
This makes sense, as the paths that the algorithm uses to find the flow are usually not the same as the paths that make up this flow \citep{ahuja1993network}. 
Step $4$ of the algorithm in Figure \ref{algBlock} describes how the individual paths are extracted from the flow.
The shortest path is found using a Dijkstra search \citep{dijkstra1959note} weighted by the length and the inverse of the width of the edges. This corresponds to finding the path of least resistance, as the fracture apertures and dynamic fluid viscosity values are constant for this study.
By iteratively adding the next path of least resistance, the algorithm adds the largest amount of fluid flow, $Q_p$, to the solution.

Edges that lie on several paths will have a flow value which is greater than the minimum value for the current path, i.e. $\beta$.
Fluid flow over such an edge contributes to several paths and is separated accordingly in this study.
In order to separate the contributions of such an edge into the contributions to the individual paths, the geometry of that edge is separated according to the $\beta$ values of those paths, as seen in Eq.~\eqref{wSplit},
\begin{linenomath*}
\begin{equation} \label{wSplit}
w_{e,p} = \dfrac{\beta}{f_e} \cdot w_e\, ,\quad  \textrm{with} \quad \sum_{p=1}^n \dfrac{\beta(p)}{f_e} = 1 ,   
\end{equation}
\end{linenomath*}
where $w_e$ is the width of this edge and $w_{e,p}$ is the width contributed by this edge to path $p$.
This ratio is used, because it is easily obtained (compared to the volume-rate-weighted average used in \citealt{Outters2000}) and because the bottlenecks from which we obtain $\beta$ are the dominant flow-determining features of any given path. 
The first path ($s$,A,D,C,$t$) of the blue flow assignment in Figure~\ref{exampleGraph}, for example, has a $\beta$-value of $\unit[2]{units}$ (see Table~\ref{pathResults}).
The first edge on this path (SA) has  $f_e = 6$ and will, therefore, contribute a width to this path of $\frac{2}{6}\cdot 6 = 2$.
This value is found for each edge for each path in the blue flow assignment in Figure~\ref{exampleGraph}, except for the vertical edge (DC) that only lies on path $s$,A,D,C,$t$.
The width contributed by this latter edge to its only path is $\frac{2}{2}\cdot 4 = 4$.

To calculate the hydraulic conductance for each path, $K_p$, the fundamental difference between combinatorial flow, $f$, and fluid flow in saturated fractures, $Q$, needs to be addressed.
The maximum combinatorial flow value that a path on a graph can accommodate is equal to the lowest edge capacity encountered on that path (Sections~\ref{sec:maxflow} and \ref{maxflow_algorithm}).
In other words, the value of $f$ on each edge of the path is reduced to the bottleneck edge value.
This is different for fluid flow in saturated fractures.
The maximum fluid flow through a fracture, or part of a fracture, is dependent on the fluid pressure gradient, $\nabla P$.
The distribution of $\nabla P$ along the path depends on the distribution of resistance to flow, offered by the fracture surfaces along that path.
In this manner, the pressure gradient over the bottleneck is maximized and, subsequently, the amount of fluid flow through this bottleneck (and therefore the entire path) is maximized as well.
To account for this discrepancy, the weighted harmonic mean (commonly used for finding the effective permeability for conductive layers in series and electrical conductance of resistors in series) of the widths, $w_{\text{harm},p}$, along the path is used (the same could be done for apertures if they vary from fracture to fracture), as seen in Eq.~\eqref{wHarm},  
\begin{linenomath*}
\begin{equation} \label{wHarm}
w_{\text{harm},p} = \left ( \sum_{e=1}^n \frac{L_e}{L_p} \cdot \frac{1}{w_{e,p}} \right )^{-1}\, ,  \quad \textrm{where} \quad L_p = \sum_{e=1}^n L_e \, ,
\end{equation}
\end{linenomath*}
where $L_p$ is the total length of $p$.
Using these obtained values, $K_p$ is calculated using the pipe approximation, described in \cite{Outters2000}.
In this approximation, a path is represented as a single fracture with a corresponding apparent width, $w_{harm}$ (length-weighted harmonic width in this case), an apparent aperture (constant in this study), and a total length, $L_p$.
$K_p$ is then calculated using the cubic law,
\begin{linenomath*}
\begin{equation} 
K_p = \dfrac{w_{\text{harm},p} a^3}{12 \mu L_p} \, .
\end{equation}
\end{linenomath*}
The paths predicted by the HSPM method are typically stair-shaped, which would cause this method to underestimate the flow rate, since most flow in a fracture occurs in a straight line between intersections.
To improve this method's performance, we introduce a correction factor, $\tau_p$, so that
\begin{linenomath*}
\begin{equation} \label{cPath}
K_p = \dfrac{w_{\text{harm},p} a^3}{12 \mu L_p}\cdot \tau_p \, , \qquad \text{where} \qquad \tau_p^{HSPM} = \dfrac{L_p} {L_{\text{short},p}} \,   \qquad \text{and} \qquad \tau_p^{ISPM} = 1 \, ,
\end{equation}
\end{linenomath*}
where $L_{\text{short},p}$ is the shortest distance between the inflow and the outflow points of $p$.
The total fluid flow, $Q$, is calculated by superpositioning the paths, while accounting for the vertical distance against gravity, $\Delta z$, resulting in 
\begin{linenomath*}
\begin{equation} \label{Qsum}
Q = \sum_{p=1}^n K_p \cdot (\Delta P - \rho g \Delta z_p)\, . % - because g = +9.81
\end{equation}
\end{linenomath*}
Eqs.~\eqref{wHarm}, \eqref{cPath}, and \eqref{Qsum}  are implemented in the function \textit{estimateVolRate}, which evaluates $Q_p$ for each path and adds it to $Q$. Figure~\ref{algBlock} summarizes the algorithm used to evaluate the produced graph networks.
%
%
%
%
%%%%%%%%%%%%%%%%%%%%%%%%%%%%%%%%%%
%%%
%%% ALGORITHM BLOCK
%%%
%%%%%%%%%%%%%%%%%%%%%%%%%%%%%%%%%%
\begin{figure}[htbp]
\begin{mdframed}

\textbf{\underline{Algorithm: graph2Q}}

\smallskip

\begin{tabular}{@{}llrlr} % @{} used to remove indent on the left
\textbf{Inputs:}& Network, 				& $G = (V,E)$; 	 &  Edge capacities, 	& $c:E$;\\
& Source and sink nodes,  & $s/t$;  		 &  Edge lengths, 		& $L:E$; 
\end{tabular}

\textbf{Outputs:}
\begin{tabular}{l}
A flow $f$ from $s$ to $t$ of maximum value, obtained by choosing successive shortest augmenting paths;\\
A shortest-path decomposition of this flow, assessing the volume rate, $Q_p$, of each path, $p$;\\
The volume rate value, $Q$, for the entire network
\end{tabular}
\medskip
\begin{itemize}[leftmargin=12pt,labelsep=0.2cm]
\item[\textbf{1}] Initialize $f_e = 0$, for all edges $e \in E$, and $Q = 0$
\item[\textbf{2}] Create residual network $G_r = (V,E)$ with residual capacities $cr_e = c_e - f_e$

\item[\textbf{3}] While $\exists$ a shortest s/t path $p$ in graph $\Big(V,\{e \in E : cr_e > 0 \}\Big)$	
	\begin{itemize}
	\item[\textbf{3.1}] Find the smallest $cr_e$ value on $p$: \hspace{10pt}
     $\alpha$ = min \{ $cr_e : e \in p$ \} 
     \item[\textbf{3.2}] For each $e \in p$:
     \vspace{2pt}
     \begin{itemize}
     \setlength\itemsep{1pt}
     \item[\textbf{3.2.1}] Add new flow to $G$: \hspace{10pt}
               $f_e \mathrel{+}= \alpha$          
     \item[\textbf{3.2.2}] Remove new flow from $G_r$: \hspace{10pt}
     $cr_e \mathrel{-}= \alpha$
     \item[\textbf{3.2.3}] Add new flow to $e'$, the reverse edge of $e$ on $G_r$: \hspace{10pt}
     $cr_{e'} \mathrel{+}= \alpha$
     \end{itemize}
	\end{itemize}

\item[\textbf{4}] While $\exists$ a shortest s/t path $p$ in graph $\Big(V,\{e \in E : f_e > 0 \}\Big)$	
\begin{itemize}
	\item[\textbf{4.1}] Find the lowest $f_e$ value on $p$: \hspace{10pt}
    $\beta$ = min \{$f_e : e \in p$\}
    
    \item[\textbf{4.2}] Reduce the flow $f_e$ by $\beta$: \hspace{10pt}
    \hspace{10pt}$\forall \hspace{1pt} e \in p$:
    $f_e \mathrel{-}= \beta$
    
    \item[\textbf{4.3}] $Q \hspace{4pt} \mathrel{+}= \text{estimateVolRate}( p, \beta )$
   
	\end{itemize}
\end{itemize}
\medskip
A detailed description of the function \textit{estimateVolRate} can be found in Section~\ref{sec:backTranslation}.
\end{mdframed}
\caption{Description of the algorithm used on the produced graph networks. Steps $2$ and $3$, which constitute the Edmonds--Karp max flow algorithm, choose the shortest paths with respect to the hop distance, though the edge length is preferable.}
\label{algBlock}
\end{figure}
%%%%%%%%%%%%%%%%%%%%%%%%%%%%

%%%%%%%%%%%%%%%%%%%%%%%%%%%%
%%%
%%% TABLE FOR PATH EXTRACTION
%%%
\begin{table}
\centering
\setlength{\tabcolsep}{12pt} % Default value: 6pt
\captionof{table}{Flow paths extracted from the blue flow assignment in Figure~\ref{exampleGraph}, as well as the results for each path obtained, using the following values: $a^3/(12\mu) = 1$, $\Delta P = 1$, $\Delta z = 0 \hspace{4pt} \forall p$
($w_{\text{harm}}$ from Eq.~\eqref{wHarm}, Q from Eq.~\eqref{Qsum}).
These results clearly show that there is no direct correlation between the calculated path flow values and the resulting $Q$ values, as all paths can accommodate $\unit[2]{units}$ of flow.
The total amount of $Q$ for this flow decomposition is $\unit[1.57]{units}$.
}
\begin{tabular}{lllll}
\hline
Path & Length & $\beta$ & $w_{\text{harm}}$ & Q	\\
\hline
$s$,A,D,C,$t$ & 4 & 2 & 2.29 & 0.57	\\
$s$,A,D,E,$t$ & 4 & 2 & 2 	 & 0.5 	\\
$s$,A,B,C,$t$ & 4 & 2 & 2    & 0.5 	\\
\hline
\end{tabular}
\label{pathResults}
\end{table}
%%%%%%%%%%%%%%%%%%%%%%%%%%%%
%
The deviation of flow rates, $\varepsilon_\text{HSPM}$ and $\varepsilon_\text{ISPM}$, between the developed methods and the DS is given by
\begin{linenomath*}
\begin{equation} \label{Qcomp}
\varepsilon_\text{HSPM} = \left ( \frac{Q_\text{HSPM}}{Q_\text{DS}} - 1 \right ) \cdot \unit[100]{\%} \,  \qquad \text{and} \qquad \varepsilon_\text{ISPM} = \left (\frac{Q_\text{ISPM}}{Q_\text{DS}} - 1\right ) \cdot \unit[100]{\%} \, .
\end{equation}
\end{linenomath*}

%%%%%%%%%%%%%%%%%%%%%%%%%%%%%%%%%%%%%%%%%%%%%%%%%%%%%%%%%%%%%%%%%
%%%
%%% NUMERICAL SIMULATIONS
%%%
\subsection{Numerical simulations}
%%%%%%%%%%%%%%%%%%%%%%%%%%%%%%%%%%%%%%%%%%%%%%%%%%%%%%%%%%%%%%%%%
This study uses GEOS, developed at Lawrence Livermore National Laboratory, for the direct simulations (DS) to verify the accuracy of the methods under development. 
GEOS is a massively parallel, multi-physics numerical framework, which was designed mainly to enable simulations of subsurface hydraulic reservoir stimulations.
It has been used to simulate hydraulic fracturing  \citep{settgast2016fully, vogler_2017a}, reactive transport \citep{walsh2013permeability}, drawdown in geothermal systems \citep{fu2016thermal}, immiscible fluid flow \citep{walsh2013fracture}, shearing of fractures using FEM \citep{annavarapu2015weighted}, crack growth using XFEM methods \citep{annavarapu_2016}, and hydro-mechanically coupled flow in rough fractures \citep{vogler_2016a,vogler_2018}. 
GEOS was chosen for DS as it offers an internal mesh generator, which greatly simplifies running simulations on a wide range of DFNs. 
Additionally, its wide range of capabilities make GEOS a suitable platform to test potential extensions of the presented method against. 

For the DS conducted here, flow in fractures is discretized and computed, employing the finite volume method (FVM) and modeled using lubrication theory and a parallel plate assumption \citep{louis1969,cameron1976basic,pantonincompressible,settgast2016fully},
\begin{linenomath*}
\begin{equation}
\frac{\partial (\rho  a ) }{\partial t} -  \frac{1}{12 \mu}  \nabla   \cdot  \left [\rho  a^3 \nabla (P + \rho g z) \right ] = 0 \, ,
\label{eq:lub}
\end{equation}
\end{linenomath*}
where $\rho$ is the fluid density, $a$ is the fracture aperture, $t$ is time, $\mu$ is the dynamic fluid viscosity, and $P$ is the fluid pressure. In this study, $\rho$, $\mu$, and $a$ are assumed to be constant.
Hence, Eq.~\eqref{eq:lub} reduces to
\begin{linenomath*}
\begin{equation}
%\nabla   \cdot \left [  a^3 \nabla (P + \rho g z) \right  ] = 0 \, .
\nabla   \cdot \nabla \left (P + \rho g z \right ) = 0 \, .
\end{equation}
\end{linenomath*}
%
%
%
%
%
%%%%%%%%%%%%%%%%%%%%%%%%%%%%%%%%%%%%%%%%%%%%%%%%%%%%%%%%%%%%%%%%%
%%%
%%% DFN GENERATION
%%%
\subsection{DFN generation} \label{sec:DFNgen}
\begin{figure}[htbp]
\centering
a) 
\includegraphics[width=0.49\textwidth]{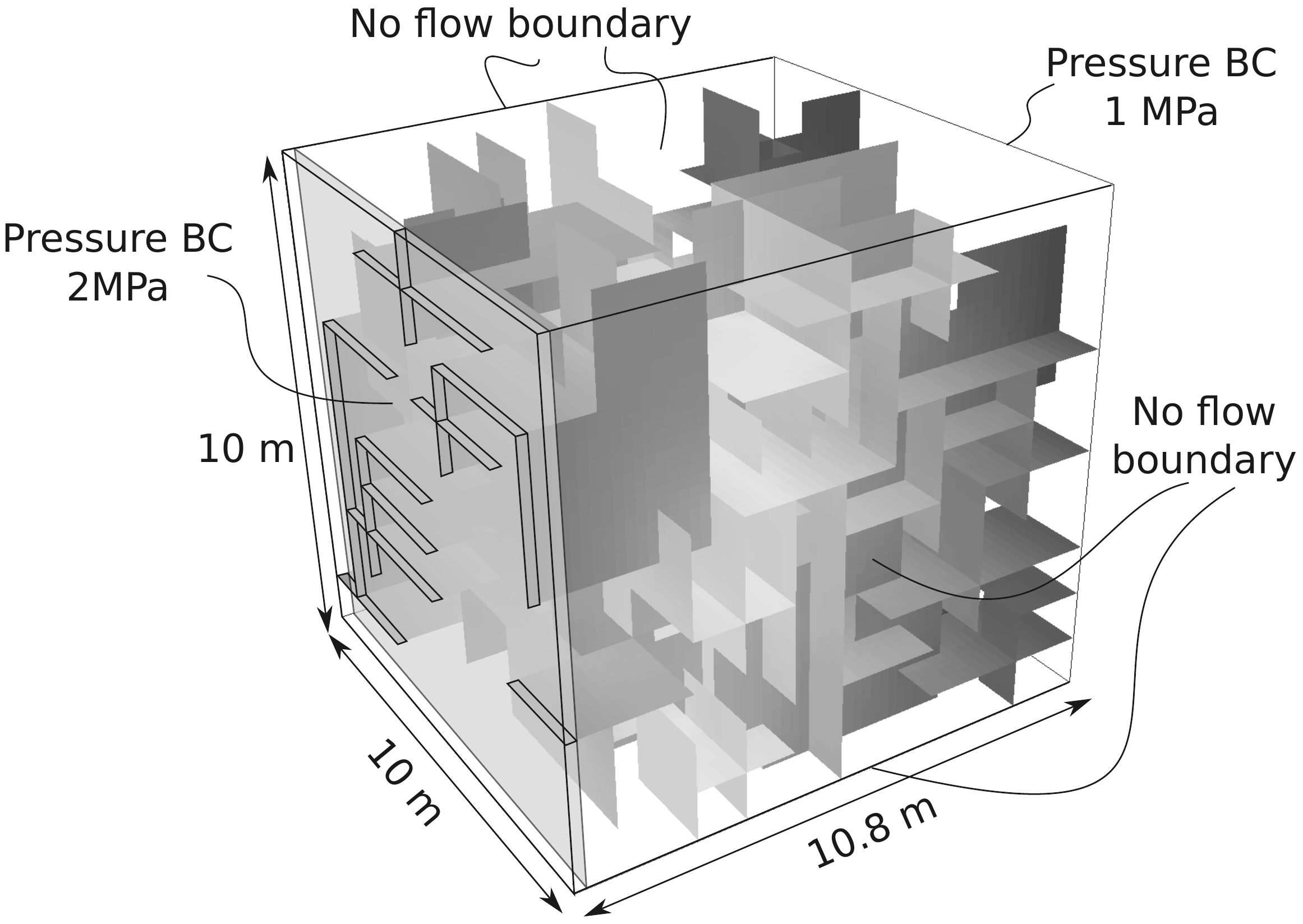}
\hfill
b) 
\includegraphics[width=0.35\textwidth]{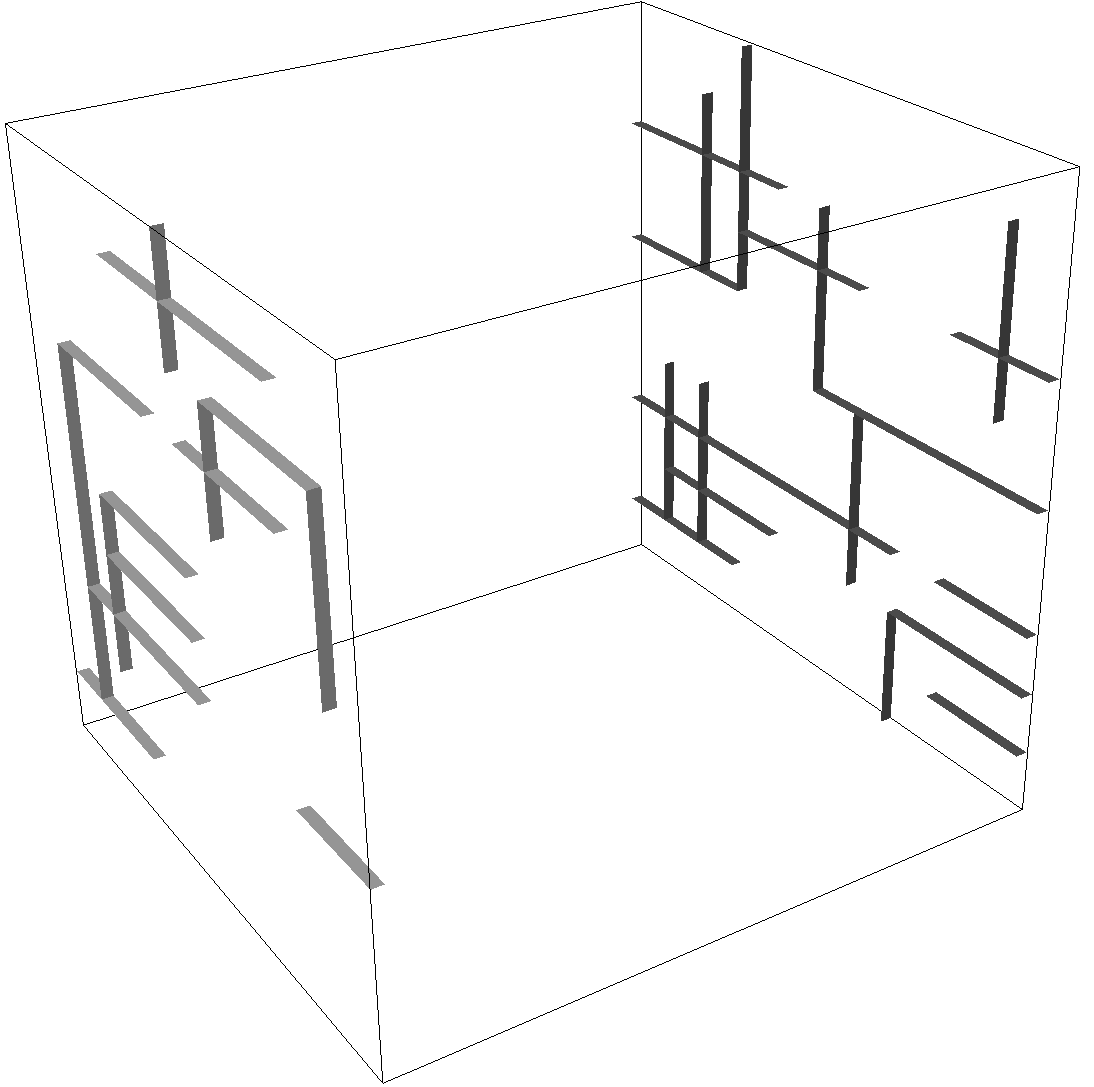}
\caption{\textbf{a)} DFN with 230 orthogonal fractures. The pressure boundaries act on the faces and, therefore, lie on $x = \unit[- 0.3]{m}$ and $x = \unit[10.3]{m}$. 
\textbf{b)} Extensions of the fractures that intersect with the $y-z$-plane boundaries of the $\unit[10]{m} \times \unit[10]{m} \times \unit[10]{m}$ volume. 
}
\label{GEOS_sim}
\end{figure}

%%%%%%%%%%%%%%%%%%%%%%%%%%%%%%%%%%%%%%%%%%%%%%%%%%%%%%%%%%%%%%%%%
The main results presented in this work are from a large amount of stochastically generated 3D DFNs with a range of fracture densities. 
A predefined amount of \unit[2.4]{m} by \unit[3.4]{m} fractures (aspect ratio of 1.42) is placed in a $\unit[10]{m} \times \unit[10]{m} \times \unit[10]{m}$ volume for each DFN.
The fracture spacing and orientation in these 3D simulations is chosen according to a randomly generated discrete uniform distribution. The orientations are perpendicular to one of the axis directions and the possible placements are every meter.
All DFNs are generated in such a way that no fracture can lie completely within a space already occupied by other fractures. 
Those fractures that intersect the $y-z$ plane boundaries (not those parallel to it), are elongated by \unit[0.4]{m}, by adding fractures outside of the boundary, to accommodate the extraction of the volume rate results.
Because pressure is defined on fracture faces in the DS, the actual pressure boundaries are, thus, located at $x = \unit[-0.3]{m}$ and $x = \unit[+10.3]{m}$.
Figure~\ref{GEOS_sim} shows a stochastically generated DFN and describes the boundary conditions used in the simulations.
Table~\ref{tab_sim_param} lists the parameters used in these simulations.
%
%
%
%
%%%%%%%%%%%%%%%%%%%%%%%%%%%%%%%%%%%%%%%%%%%%%%%%%%%%%%%%%%%%%%%%%
%%%
%%% FIGURE - BOUNDARY CONDITION
%%%
%%%%%%%%%%%%%%%%%%%%%%%%%%%%%%%%%%%%%%%%%%%%%%%%%%%%%%%%%%%%%%%%%
%
%
%
%
%%%%%%%%%%%%%%%%%%%%%%%%%%%%%%%%%%%%%%%%%%%%%%%%%%%%%%%%%%%%%%%%%
%%%
%%% MESH REFINEMENT
%%%
\subsection{Mesh refinement}
%%%%%%%%%%%%%%%%%%%%%%%%%%%%%%%%%%%%%%%%%%%%%%%%%%%%%%%%%%%%%%%%%
%
\cite{Mourzenko2011} and \cite{ebigbo2016inclusion} show that the mesh discretization length of a DS has an influence on the resulting permeability. 
In order to perform the comparison described in Eq.~\eqref{Qcomp}, a systematic mesh refinement is conducted for several of the stochastic DFNs.
The following equation, as proposed by \cite{Mourzenko2011}, shows how the actual volume rate result, $Q_\text{DS}$, depends on the calculated result, Q$_{\delta}$, with discretization length, $\delta$, average fracture length, $l$, and a refinement coefficient, $D$,
\begin{linenomath*}
\begin{equation}
Q = \dfrac{Q_{\delta}}{1+D \dfrac{\delta}{l}} \, .
\end{equation}
\end{linenomath*} 
This linear behavior is explained by \cite{Mourzenko2011} as a natural consequence of using a first-order discretization of the equations of flow.
They investigated the refinement coefficient, $D$, for fractures of different shapes, including rectangles with an aspect ratio of 6 
and found a dependency of the $D$-values on this shape and on the dimensionless fracture density, among other factors. 

This study's $D$-values are obtained systematically for the stochastic DFNs using meshes with $\delta$s between $\unit[0.05]{m}$ and $\unit[0.5]{m}$.
One simulation each is chosen from cases with 150, 190, 230, 270, and 310 fractures.
The $D$-values for the other cases are interpolated from the obtained values. 
%
%
%
%
%
%%%%%%%%%%%%%%%%%%%%%%%%%%%%%%%%%%%%%%%%%%%%%%%%%%%%%%%%%%%%%%%%%
%%%
%%% VERIFICATION OF HSPM APPROACH
%%%
\section{Verification of the HSPM and ISPM methods} \label{sec:results}
%%%%%%%%%%%%%%%%%%%%%%%%%%%%%%%%%%%%%%%%%%%%%%%%%%%%%%%%%%%%%%%%%
%
%
The main results of this proof-of-concept investigation are presented in this section.
First, the results of both methods for simple test cases are compared to the reference DS results.
Next, the same comparison is presented for 100 stochastic DFNs with a range of fracture densities.
Lastly, an assessment of the uncertainty with respect to fracture density and fracture constellation for 1000 stochastic DFNs is conducted to show how the presented methods can be used to quantify uncertainty in flow rate for a DFN parameter space.
%
%
%
%
%%%%%%%%%%%%%%%%%%%%%%%%%%%%%%%%%%%%%%%%%%%%%%%%%%%%%%%%%%%%%%%%%
%%%
%%% SIMPLE TEST CASES
%%%
\subsection{Simple test cases}\label{sec:test}
\begin{figure}[htbp]
\centering
\includegraphics[width=\textwidth]{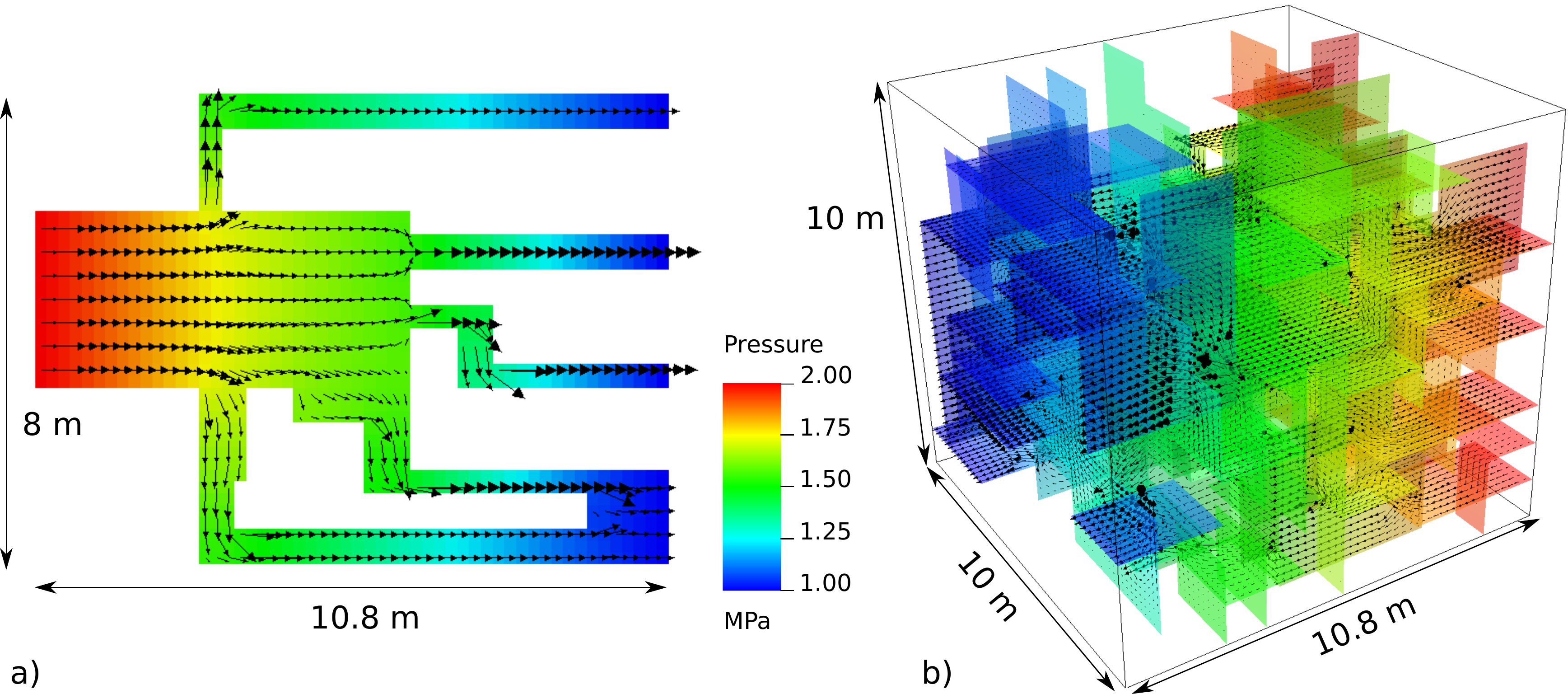}
\caption{ Two test-case direct simulations. \textbf{a)} a 2D case, \textbf{b)} a stochastic 3D case with 230 fractures of \unit[2.4]{m} by \unit[3.4]{m}. Arrows show the fluid velocity and are scaled according to the fluid-velocity magnitude. Pressure is shown with color contours.}
\label{benchmarks}
\end{figure}

\begin{figure}[htbp]
%%%%%%%%%%%%%%%%%%%%%%%%%%%%%%%%%%%%%%%%%%%%%%%%%%%%%%%%%%%%%%%%%
%%%
%%% FIGURE - TEST CASE GRAPHS
%%%
%%%
\centering
a) 
\includegraphics[width=0.54\textwidth]{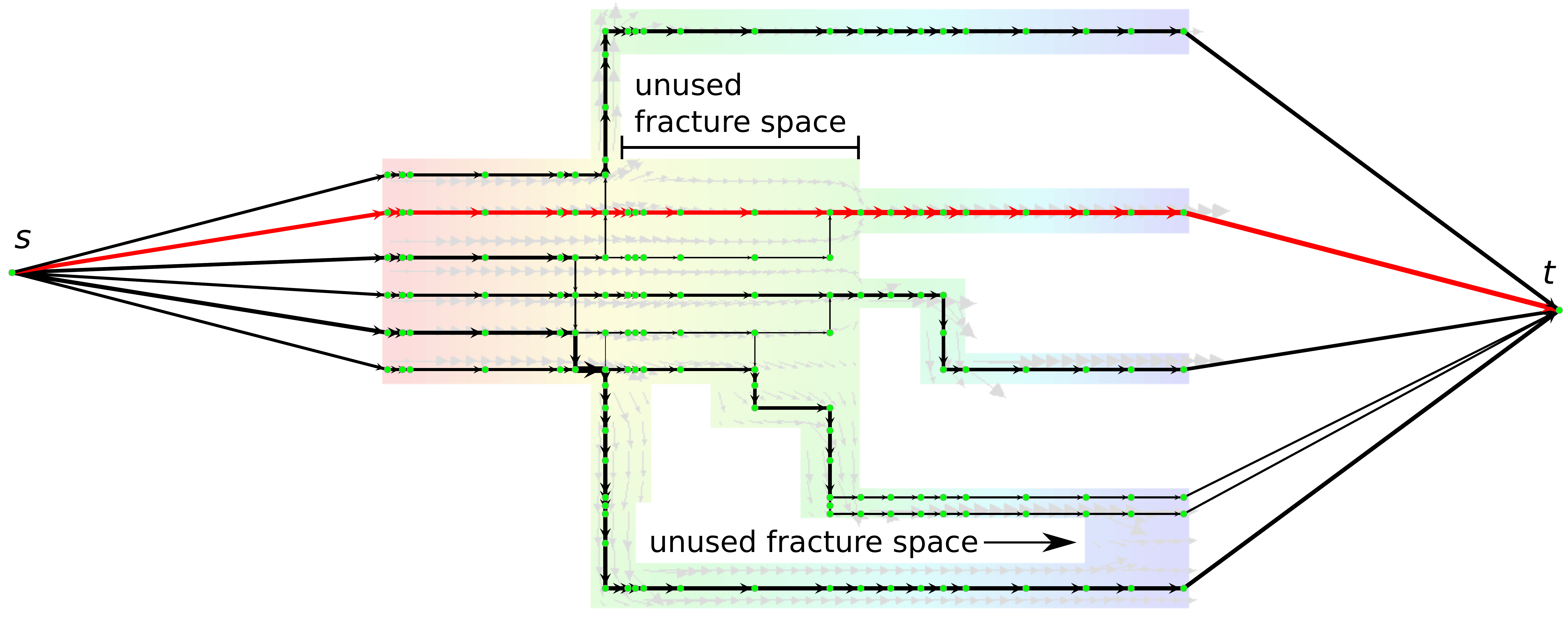}
\hfill 
b) 
\includegraphics[width=0.36\textwidth]{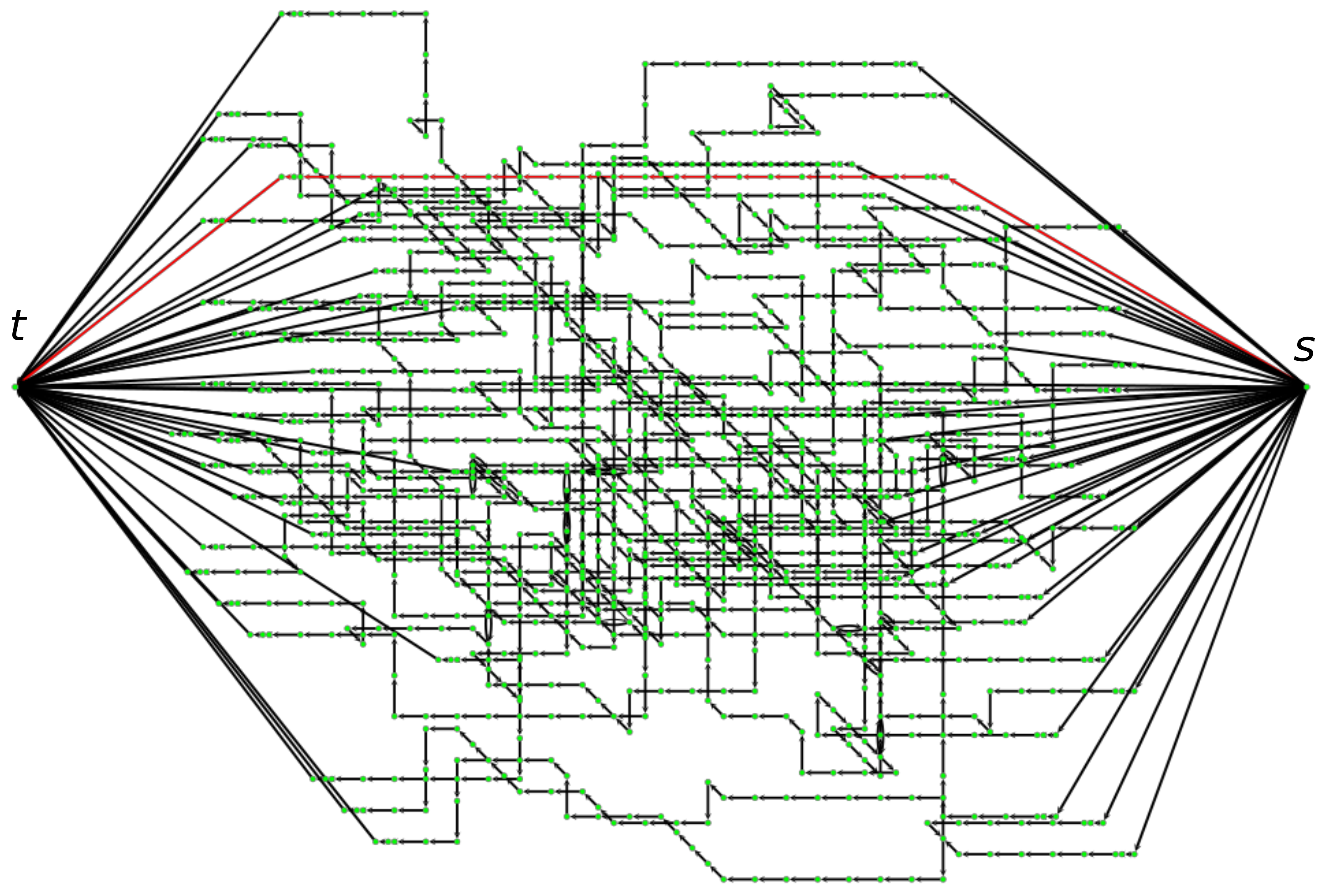}
\caption{The Edmonds-Karp result of the graphs created by the HSPM method for the DFNs shown in Figure \ref{benchmarks}. 
The red path is the first path found by the algorithm. 
Edges and vertices (green circles), where no flow occurs in the Edmonds-Karp solution, have been removed.
\textbf{a)} The graph is overlain on the pressure result from Figure \ref{benchmarks}a.
The super-source, $s$, and super-sink, $t$, are connected by 13 paths.
This solution does not cover all the fracture space in the DFN, as can be seen in the central fracture just above the red path and in the bottom right corner.
\textbf{b)} $s$ and $t$ are connected by 40 paths.
}
\label{testGraphs}
\end{figure}

%%%%%%%%%%%%%%%%%%%%%%%%%%%%%%%%%%%%%%%%%%%%%%%%%%%%%%%%%%%%%%%%%
Figure~\ref{benchmarks} shows two of the many simulations that were used to test the developed methods: a 2D simulation with five paths that originate in a single fracture and a 3D simulation with 230 fractures generated as described in Section~\ref{sec:DFNgen}.
Figure~\ref{testGraphs} shows the Edmonds--Karp flow result on the graphs obtained from the two test cases in Figure~\ref{benchmarks}, employing HSPM. 
These graphs do not correspond to the full graphs extracted from the DFN, as edges and vertices without flow have been removed to improve visibility. 
%
%
%
%%%%%%%%%%%%%%%%%%%%%%%%%%%%%%%%%%%%%%%%%%%%%%%%%%%%%%%%%%%%%%%%%
%%%
%%% FIGURE - 2 SIMPLE TEST CASES
%%%
%%%
%%%%%%%%%%%%%%%%%%%%%%%%%%%%%%%%%%%%%%%%%%%%%%%%%%%%%%%%%%%%%%%%%
%
%
%%%%%%%%%%%%%%%%%%%%%%%%%%%%%%%%%%%%%%%%%%%%%%%%%%%%%%%%%%%%%%%%%
%
%
%
%
Table \ref{benchmark_results} shows the $Q$, $\varepsilon_\text{HSPM}$, and $\varepsilon_\text{ISPM}$ values, obtained for these two test cases, as well as the amount of nodes and edges that were produced. 
It also shows the time it takes to run the three methods (DS, HSPM, ISPM).
The computational costs for the DS are calculated using a \unit[2.6]{GHz} XEON E5-2670 server processor and  do not include extracting the $Q$-values using a Python interface (another \unit[30]{s} per simulation). 
The computational costs for the HSPM and ISPM methods are calculated using a \unit[2.9]{GHz} I7-3520M laptop processor. These costs include the extraction of the graph, getting the result from the Edmonds-Karp algorithm, extracting the paths, and calculating the final flow rate result, but do not include the importation of the needed Python modules, which occurs only once, irrespective of the amount of investigated DFN in a loop.
%
%
%
%
%%%%%%%%%%%%%%%%%%%%%%%%%%%%%%%%%%%%%%%%%%%%%%%%%%%%%%%%%%%%%%%%%
%%%
%%% TABLE - TEST CASE RESULTS
%%%
\begin{table}[htbp]
\caption{Simple test case results for DS, HSPM, and ISPM methods. Shown are flow rates, $Q$, differences between flow rates ($\varepsilon_\text{HSPM} = \left [ Q_\text{HSPM}/Q_\text{Ds} - 1 \right ] \cdot \unit[100]{\%}$, $\varepsilon_\text{ISPM} = \left [ Q_\text{ISPM}/Q_\text{Ds} -1 \right ] \cdot \unit[100]{\%}$), computational cost, CPU, number of vertices, $v$ (nodes in the DS case), number of edges, $\#e$, number of elements, $el$, and number of faces, $F$  (only DS uses $el$ and $F$).}
\label{benchmark_results}
\begin{center}
\begingroup
\setlength{\tabcolsep}{20pt} 
\small
\begin{tabular}{lcrr}
\toprule
& Units 			& 2D      	& 3D      	\\ \midrule
$Q_\text{DS}$    	& $\unit{mL/s}$ 	& 0.0218 	& 0.302 	\\
$Q_\text{HSPM}$  	& $\unit{mL/s}$	& 0.0216 	& 0.300 	\\
$Q_\text{ISPM}$ 		& $\unit{mL/s}$ 	& 0.0110 	& 0.306 	\\
$\varepsilon_\text{HSPM}$   	& $\unit{\%}$		& -0.82   	& -0.57   	\\
$\varepsilon_\text{ISPM}$   	& \%		    	& -49.69  	& +1.52   	\\
$\text{CPU}_\text{DS}$     	& $\unit{s}$     	& 14.00    	& 630.00    \\
$\text{CPU}_\text{HSPM}$   	& $\unit{s}$     	& 0.25    	& 3.53    	\\
$\text{CPU}_\text{ISPM}$   	& $\unit{s}$     	& 0.03    	& 1.96    	\\
$v_\text{HSPM}$     	& --	      			& 245     	& 4\,393    	\\
$v_\text{ISPM}$     	& --      			& 26      	& 1\,495    	\\
$v_\text{DS}$	   		& -- 			    & 14\,025    	& 143\,055  	\\
$\#e_\text{HSPM}$     	& --      			& 670     	& 14\,516   	\\
$\#e_\text{ISPM}$     	& --     	   		& 58      	& 5\,478    	\\
$el_\text{DS}$      	& --      			& 10\,800    	& 135\,000  	\\
$F_\text{DS}$       	& --     			& 35\,516    	& 412\,900  	\\
\bottomrule
\end{tabular}
\endgroup
\end{center}
\end{table}
%%%%%%%%%%%%%%%%%%%%%%%%%%%%%%%%%%%%%%%%%%%%%%%%%%%%%%%%%%%%%%%%%
%
%
%
%
The HSPM method does very well for both test cases (Figure~\ref{benchmarks}).
Both methods are orders of magnitude faster than the DS, with the ISPM method being faster than the HSPM method.
The ISPM method does not perform well in 2D, as the assumption that flow occurs mainly along the longest dimension of the fracture is easily violated in 2D.
The 2D case presented here, for example, can be resolved much better ($\varepsilon_{\text{ISPM}} = $ \unit[-7.77]{\%}), when we use the width perpendicular to the fluid flow direction for each edge.
Surprisingly, the assumptions of the ISPM method work very well for most 3D cases with a large fracture density.
In contrast, when the width perpendicular to flow is used in the 3D ISPM method, the flow rate results are more than double of those from the DS in most cases.
We investigate additional 3D cases in the following section.
%
%
%
%
%%%%%%%%%%%%%%%%%%%%%%%%%%%%%%%%%%%%%%%%%%%%%%%%%%%%%%%%%%%%%%%%%
%%%
%%% GRAPH THEORY RESULTS COMPARISON TO REAL DFN
%%%
\subsection{Investigation of 100 stochastic DFNs} \label{test}
\begin{figure}[htbp]
\centering
a)\includegraphics[width=0.47\textwidth]{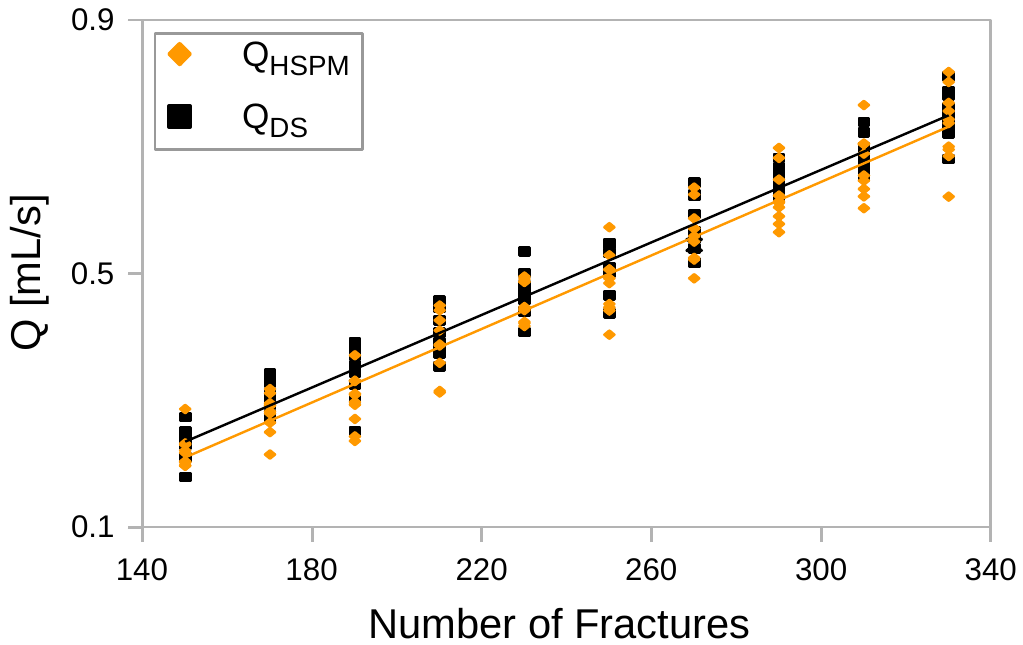}
\hfill
b)\includegraphics[width=0.47\textwidth]{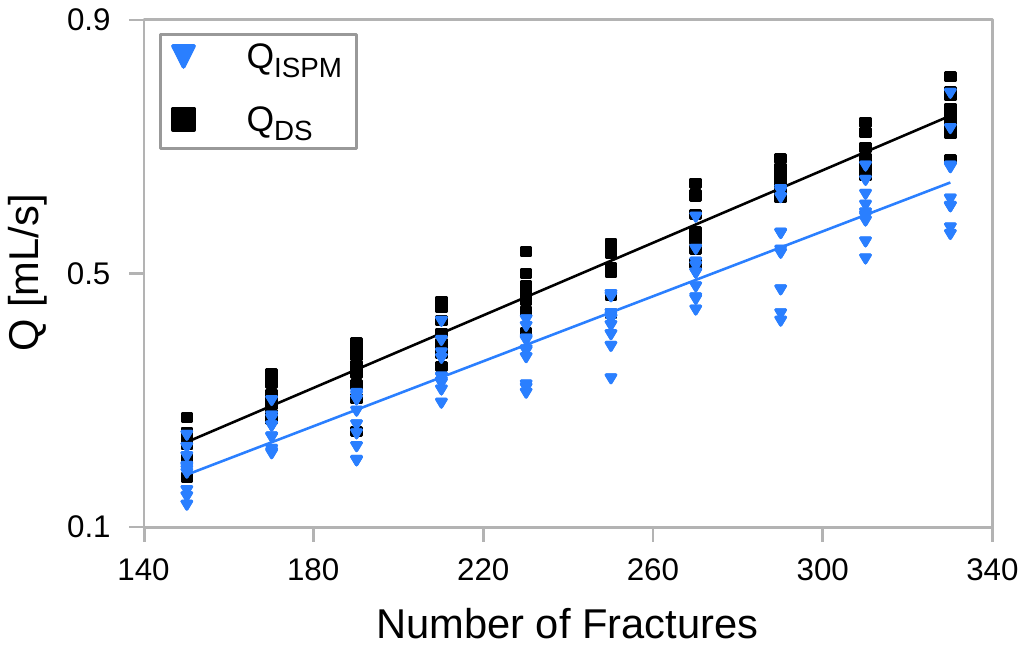}
\caption{Volumetric flow rate, $Q$, results of the two developed methods and DS for 100 stochastic simulations for a range of fracture densities.
Each chosen fracture density has ten simulations. 
A least-squares fit trend line is depicted for each method.
}
\label{fig:Q_comparison_seg}
\end{figure}

\begin{figure}[htbp]
\centering
a)\includegraphics[width=0.47\textwidth]{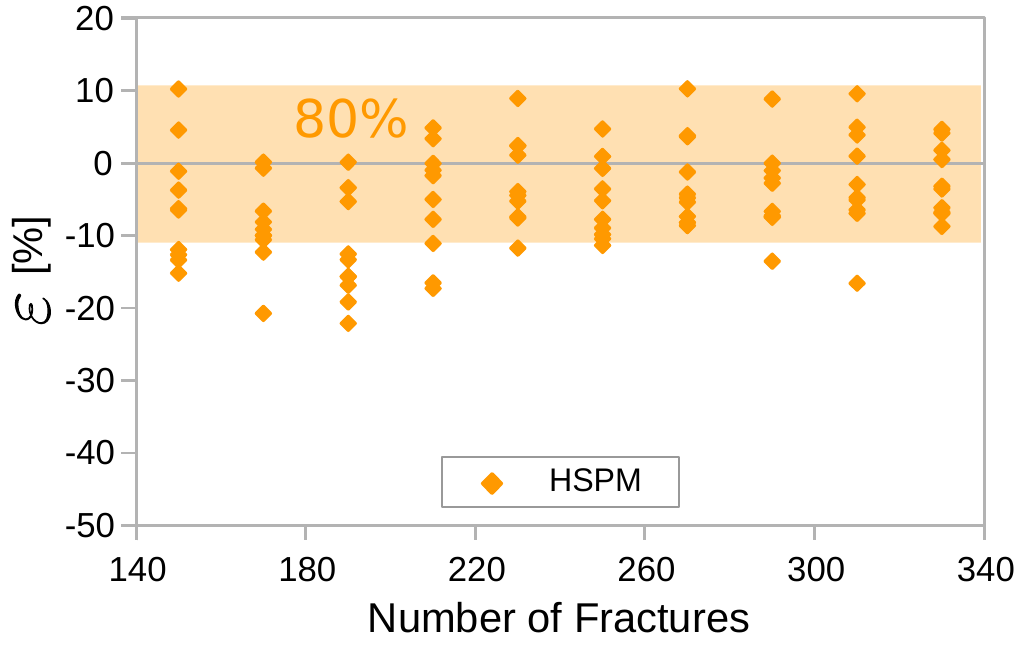}
b)\includegraphics[width=0.47\textwidth]{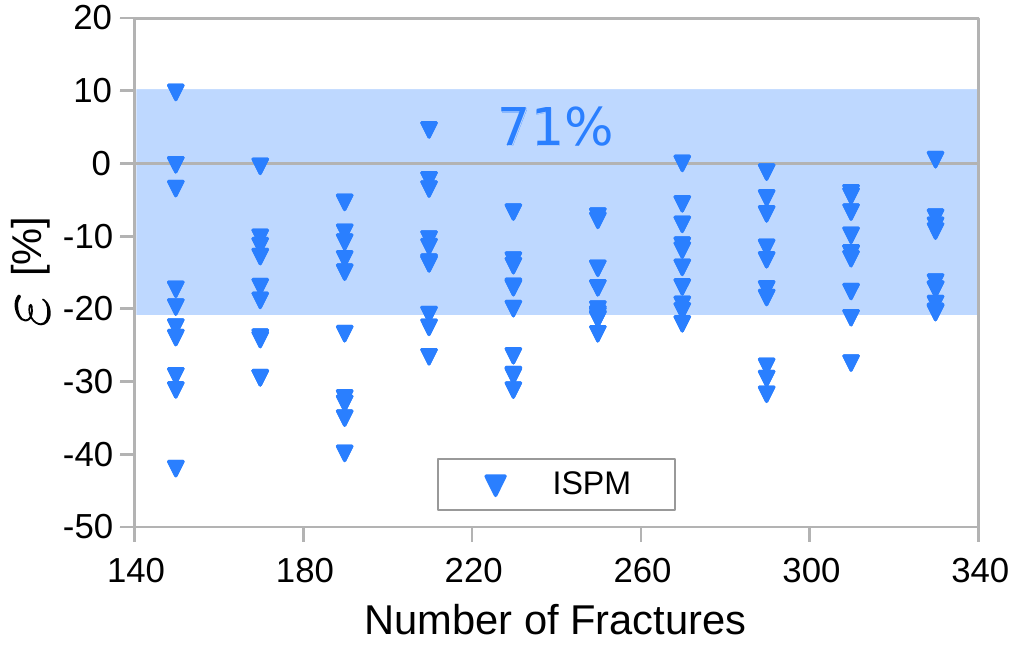}
\caption{Differences, $\varepsilon$, in predicted volumetric flow rate, $Q$, in \unit{\%} for \textbf{a)} the HSPM method and  \textbf{b)} the ISPM method, compared to the reference DS results for 100 simulation test cases with stochastic DFNs, covering a range of fracture densities.
$\epsilon =$ \unit[+3]{\%}, corresponds to an overestimation of $Q$ by \unit[3]{\%}, compared to the DS results.
$\unit[80]{\%}$ of the $Q$ results predicted by the HSPM method lie between $\unit[+10.3]{\%}$ and $\unit[-10.6]{\%}$.
$\unit[71]{\%}$ of the $Q$ results of the ISPM method lie between $\unit[+9.7]{\%}$ and $\unit[-21.0]{\%}$.
}
\label{accuracy}
\end{figure}

\begin{figure}[htbp]
\centering
a)\includegraphics[width=0.47\textwidth]{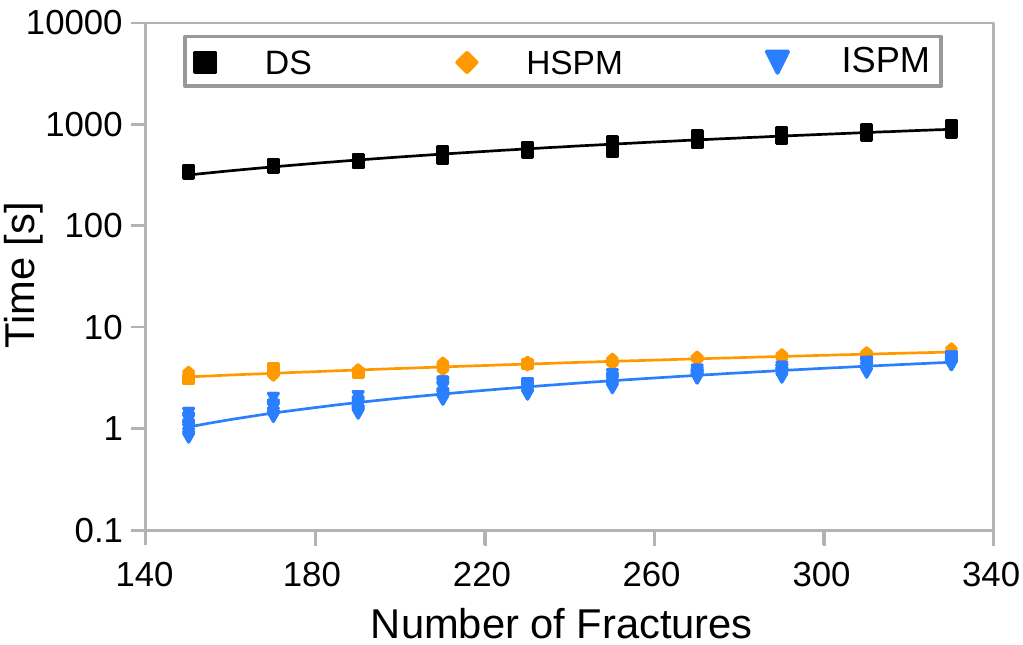}
\hfill
b)\includegraphics[width=0.47\textwidth]{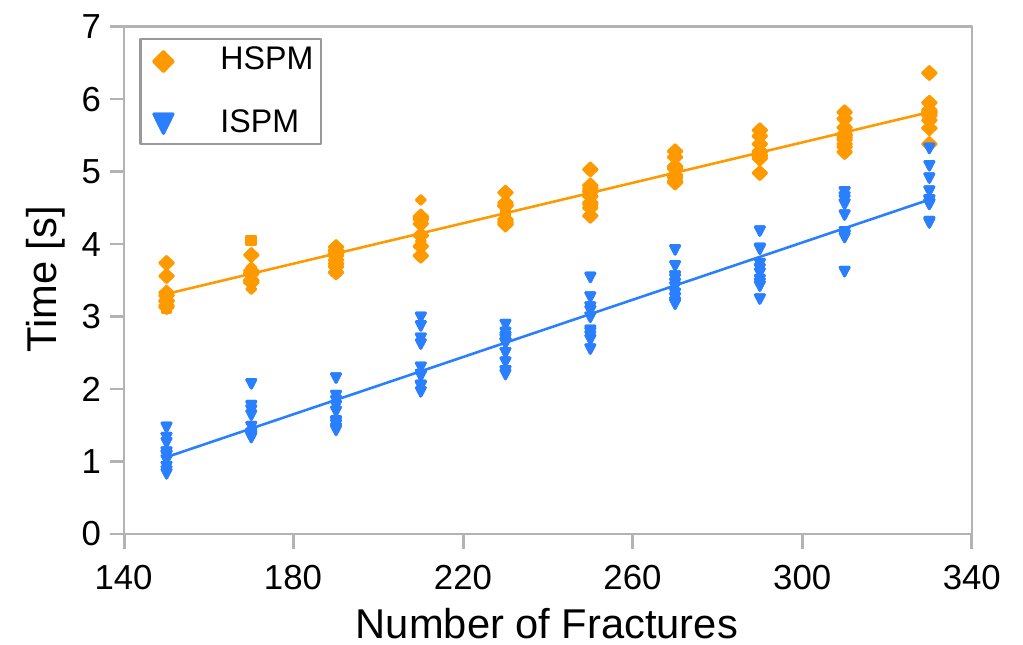}
\caption{Computational cost comparison of \textbf{a)} the DS, HSPM, and ISPM methods and \textbf{b)} the HSPM and ISPM methods for 100 stochastic DFNs with a range of fracture densities.
A least-squares fit trend line is depicted for each method.
}
\label{Comp_cost_100_cases}
\end{figure}

\begin{figure}[htbp]
\centering
a)\includegraphics[width=0.47\textwidth]{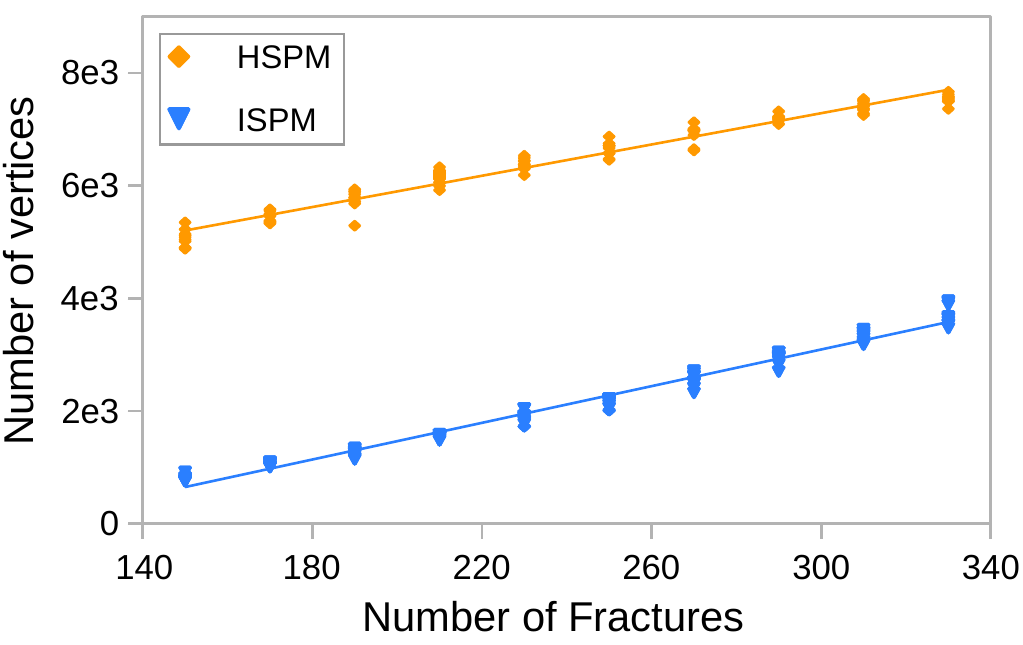}
b)\includegraphics[width=0.47\textwidth]{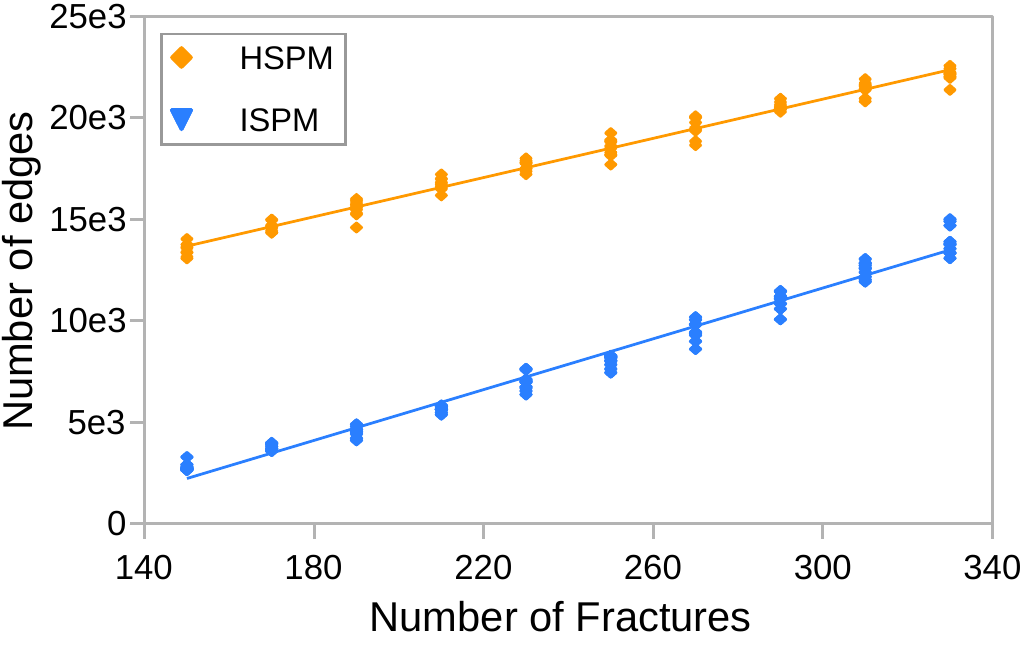}
\caption{
Comparison of \textbf{a)} the number of vertices and \textbf{b)} the number of edges created by the HSPM method and the ISPM method for 100 simulation test cases each, with stochastic DFNs with a range of fracture densities (from Figure \ref{fig:Q_comparison_seg}). 
}
\label{vert_edge}
\end{figure}

%%%%%%%%%%%%%%%%%%%%%%%%%%%%%%%%%%%%%%%%%%%%%%%%%%%%%%%%%%%%%%%%%
100 stochastic DFNs with a range of fracture densities were generated as discussed in Section~\ref{sec:DFNgen}. 
Each chosen fracture density has ten realizations. 
Figures~\ref{fig:Q_comparison_seg}--\ref{vert_edge} show the results of this 100-case investigation.

The range of volumetric flow rates predicted by the HSPM method is very similar to the range obtained from the DS, but values do lie slightly below those for the DS method (Figure~\ref{fig:Q_comparison_seg}a). 
The least-squares fit trend lines are parallel and almost coincide. 
For certain fracture densities, the HSPM method predicts a larger range than predicted by the DS (e.g. 290 fractures). 
However, the opposite phenomenon also occurs (e.g. 230 fractures). 
The range of the ISPM method's results (Figure~\ref{fig:Q_comparison_seg}b) show a greater fluctuation.
Both small ranges and large ranges occur (170 and 290 fractures, respectively). 
The least-squares fit trend line of the ISPM method lies below and slightly diverges from that of the DS with increasing fracture number.
%
%
%
%
%%%%%%%%%%%%%%%%%%%%%%%%%%%%%%%%%%%%%%%%%%%%%%%%%%%%%%%%%%%%%%%%%
%%%
%%% FIGURE - VOLUMERATE COMPARISON 
%%%
%%%%%%%%%%%%%%%%%%%%%%%%%%%%%%%%%%%%%%%%%%%%%%%%%%%%%%%%%%%%%%%%%
%
%
%
%

The volumetric flow rates predicted by the two methods are compared to the DS results in Figure~\ref{accuracy}.
The average discrepancy of the HSPM method is $\varepsilon_{\text{HSPM}} = \unit[-5.03]{\%}$.
$\unit[80]{\%}$ of the results predicted by the HSPM method lie within the $\unit[+10.3]{\%}$ and $\unit[-10.6]{\%}$ range, compared to the DS results, with more than half of the results lying within $\unit[\pm 5]{\%}$. 
No systematic error seems to occur for HSPM.

The ISPM method is less accurate (Figure~\ref{accuracy}) with an average $\varepsilon_{\text{ISPM}}$ of $\unit[-16.38]{\%}$.
$\unit[71]{\%}$ of the results of the ISPM method lie between $\unit[+9.7]{\%}$ and $\unit[-21]{\%}$ of the results of the DS, with $\unit[26]{\%}$ of the results being within $\unit[\pm 10]{\%}$.
Therefore, the ISPM method seems to systematically underestimate the DS results.
%
%
%
%
%%%%%%%%%%%%%%%%%%%%%%%%%%%%%%%%%%%%%%%%%%%%%%%%%%%%%%%%%%%%%%%%%
%%%
%%% FIGURE - ACCURACY COMPARISON 
%%%
%%%%%%%%%%%%%%%%%%%%%%%%%%%%%%%%%%%%%%%%%%%%%%%%%%%%%%%%%%%%%%%%%
%
%
%
%

Figure~\ref{Comp_cost_100_cases} shows the computational costs of the HSPM, ISPM and DS methods for the 100 stochastic DFNs.
The computational cost (explained in Section~\ref{test}) of the HSPM and ISPM methods (on a \unit[2.9]{GHz} I7-3520M laptop processor) are orders of magnitude lower, compared to the DS (on a \unit[2.6]{GHz} XEON E5-2670 server processor).
All three methods show a linear increase in this cost with increasing number of fractures (not shown for the DS). 
On average, the DS take 132 times longer for these calculations than the HSPM method.
This value is 233 when comparing the DS with the ISPM method and 1.8 when comparing the HSPM method with the ISPM method, though the HSPM method scales better with increasing fracture number. 

%
%
%
%
%
%%%%%%%%%%%%%%%%%%%%%%%%%%%%%%%%%%%%%%%%%%%%%%%%%%%%%%%%%%%%%%%%%
%%%
%%% FIGURE - COMPUTATIONAL COST COMPARISON
%%%
%%%%%%%%%%%%%%%%%%%%%%%%%%%%%%%%%%%%%%%%%%%%%%%%%%%%%%%%%%%%%%%%%
%
%
%
%
The produced number of vertices and edges increases linearly with increasing fracture density for both the HSPM and the ISPM methods (Figure~\ref{vert_edge}) and deviates little from the average for each sampled fracture density.
The DS discretization is always the same: 135'000 elements, 143'055 nodes, and 412'900 faces. 
%
%
%
%
%
%%%%%%%%%%%%%%%%%%%%%%%%%%%%%%%%%%%%%%%%%%%%%%%%%%%%%%%%%%%%%%%%%
%%%
%%% FIGURE - VERTEX AND EDGE COMPARISON
%%%
%%%%%%%%%%%%%%%%%%%%%%%%%%%%%%%%%%%%%%%%%%%%%%%%%%%%%%%%%%%%%%%%%
%
%
%
%
%%%%%%%%%%%%%%%%%%%%%%%%%%%%%%%%%%%%%%%%%%%%%%%%%%%%%%%%%%%%%%%%%
%%%
%%% APPLICATION TO A SENSITIVITY ANALYSIS
%%%
\subsection{Application to a sensitivity analysis}
\begin{figure}[htbp]
\centering
a)\includegraphics[width=0.47\textwidth]{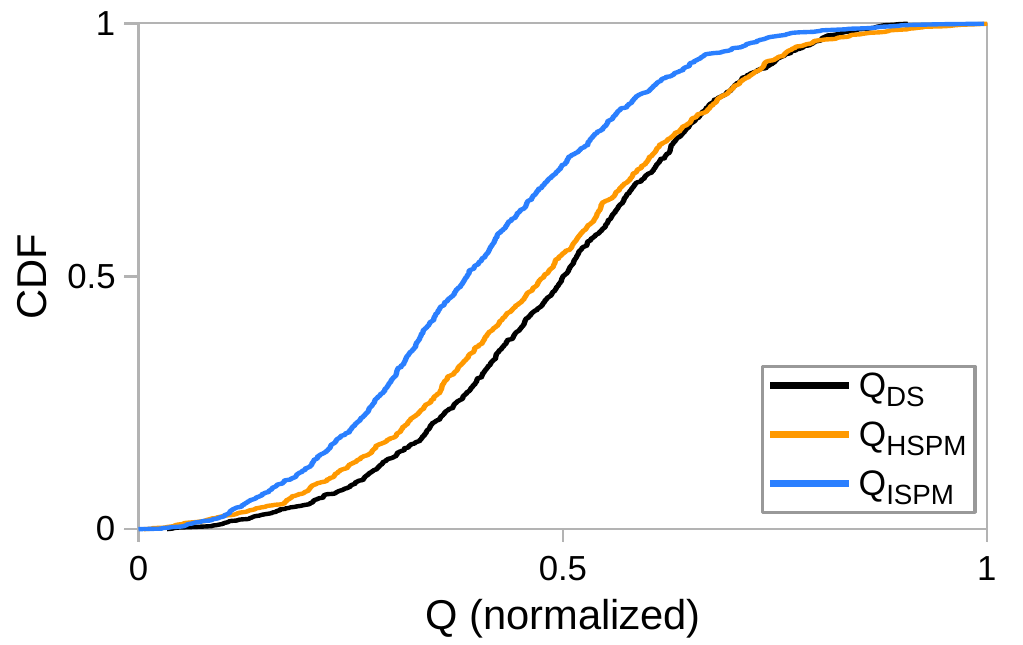}
\hfill
b)\includegraphics[width=0.47\textwidth]{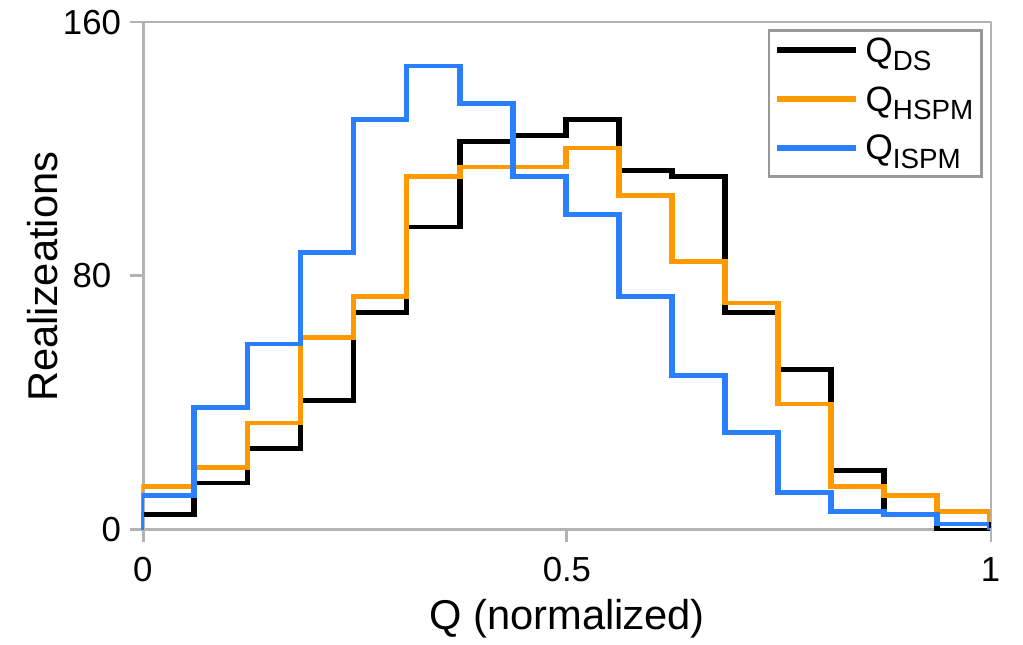}
\caption{\textbf{a)} Cumulative density functions (CDF) for normalized volumetric flow rates, $Q$, for 1000 discrete fracture networks (DFNs) with a discrete normal distribution of fracture densities with a mean of 250 and a standard deviation of 50 fractures. 
$Q$ was normalized by subtracting the smallest value and subsequently dividing through by the new maximum value.
\textbf{b)} Corresponding histograms.} 
\label{CDF}
\end{figure}

To show how the presented methods can be used in an uncertainty quantification for a DFN parameter space, an additional investigation, employing 1000 stochastic DFN realizations, is performed.
These DFNs were generated as described in Section~\ref{sec:DFNgen}, employing a discrete normal distribution of fracture densities with a mean of 250 fractures and a standard deviation of 50 fractures. 
This results in a large number of different fracture constellations, enabling us to study their effect on fluid flow rates, $Q$.
The main result of this investigation is shown in Figure~\ref{CDF}, an uncertainty analysis of a cumulative density function (CDF) of flow rate and associated histograms, based on the distribution of fracture density and associated fracture constellations.
The computational cost (as described in Section~\ref{test}) for these 1000 cases was 69 and 43 minutes, respectively, for the HSPM and ISPM calculations on a \unit[2.9]{GHz} I7-3520M laptop processor and 7.59 days for the DS on a \unit[2.6]{GHz} XEON E5-2670 server processor. 
$Q$ is normalized here ($Q'$) by subtracting the smallest value and subsequently dividing through by the new maximum value.
The CDF of the HSPM method is almost identical to that for the DS results.
The results of the ISPM method have a steeper slope,
causing the CDF to appear shifted to the left (lower $Q$-values), though the same absolute range is obtained.
The histograms show a similar trend, with the HSPM results coming very close to the DS results and the ISPM results having a slight shift to the left and higher probabilities for $Q'$ values between 0.2 and 0.4.
A confidence interval for these values can be chosen according to the needs of the specific investigation. For example, a geothermal energy production project could be interested in the minimum value needed to ensure economic operation conditions, whereas nuclear waste storage projects are targeting to stay below a certain maximum flow rate. 

%
%
%
%
%%%%%%%%%%%%%%%%%%%%%%%%%%%%%%%%%%%%%%%%%%%%%%%%%%%%%%%%%%%%%%%%%
%%%
%%% FIGURE - CDF AND HISTOGRAM
%%%
%%%%%%%%%%%%%%%%%%%%%%%%%%%%%%%%%%%%%%%%%%%%%%%%%%%%%%%%%%%%%%%%%
%
%
%
%
%%%%%%%%%%%%%%%%%%%%%%%%%%%%%%%%%%%%%%%%%%%%%%%%%%%%%%%%%%%%%%%%%
%%%
%%% DISCUSSION
%%%
\section{Discussion} \label{sec:discussion}
%%%%%%%%%%%%%%%%%%%%%%%%%%%%%%%%%%%%%%%%%%%%%%%%%%%%%%%%%%%%%%%%%
%%%
%%% OVERVIEW
%%%
The presented proof-of-concept investigation employs the Edmonds-Karp algorithm, 
among other combinatorial algorithms, to quickly compute volumetric flow rates through discrete fracture networks (DFNs).
The combinatorial optimization approach does not require traditional meshing and can accommodate complex DFN geometries, as information about fracture segment/intersection connections and occurring flow between them is sufficient to construct a graph. 
The presented approach therefore only relies on the chosen programming environment (currently Python along with the Python graph-tool \citealt{peixoto-graph-tool-2014}), and can be used on any computing platform without the need for additional software and especially meshing of DFNs, which is notoriously difficult \citep{Saevik2013}.

The Hanan Shortest Path Maxflow (HSPM) method represents fracture segments as nodes, while the Intersection Shortest Path Maxflow (ISPM) method uses fracture intersections. 
While this study employs orthogonal fractures, which are relatively common in geological systems \citep{bock1980fundamentale,dunne1990orthogonal,bai2002orthogonal}, this does not reflect a limitation of the presented approach, but merely simplifies DFN generation and meshing.

%%%
%%% SUBSECTION - ACCURACY OF THE 100 STOCHASTIC DFN
%%%
\subsection{Accuracy of the 100 stochastic DFN} 
The HSPM method exhibits a high level of accuracy for the investigation of 100 stochastic DFN test cases (Figure~\ref{accuracy}), with an average value of $\varepsilon_{\text{HSPM}}$ of \unit[-5.03]{\%}. 
This shows the applicability of this method, especially when compared to the discretization error of the DS method, which shows a linear trend between \unit[+3.4]{\%} (150 fractures) and \unit[+8.1]{\%} (330 fractures) for the 100 stochastic DFNs for $\delta = \unit[0.2]{m}$.
Though further investigation may improve upon these results, this level of accuracy is sufficient for screening purposes and is still acceptable when compared to the inherent difficulties associated with the characterization of fractured rock reservoirs at depth \citep{Faybishenko2000,adler2013fractured,wilson2015developing}. 

The ISPM method has a lower level of accuracy (Figure~\ref{accuracy}), with an  average value of $\varepsilon_{\text{ISPM}}$ of \unit[-16.38]{\%}.
\unit[71]{\%} of the results lie within the accuracy range obtained by the HSPM method. 
The other $\varepsilon_{\text{ISPM}}$ values lie below the HSPM's accuracy range.

Both methods exhibit the same, fundamental shortcoming.
Due to the difference between fluid flow in fractures and calculations done on a graph, the Edmonds-Karp algorithm tends to ignore parts of fractures, or even parts of paths, which sometimes leads to an underestimation of the pressure gradient at bottlenecks, and hence to an underestimation of the volumetric fluid flow rate, $Q$, as shown in Figures~\ref{exampleGraph} and \ref{testGraphs}a.
The order in which the HSPM method defines nodes is chosen to minimize this behavior. 

The ISPM method exhibits additional shortcomings that explain the larger range of flow rate deviations from the DS results.
An overestimation of the flow rate, $Q$, can occur when fracture intersections limit flow, since the ISPM method ignores the size of intersections between fractures.
An underestimation of $Q$ by the ISPM method can occur for multiple reasons.
Firstly, the width value attributed to an edge tends to be underestimated, where fluid flow does not occur along the longest fracture dimension, causing a reduction of the calculated flow rate for all paths which use such edges. 
Secondly, connecting the vertices along the longest dimension of the current fracture can cause an unwanted increase in path length, because the path undulates between intersections, instead of taking a more direct route.
This limitation is not easy to adjust for, as it is impossible to predict flow paths beforehand.
Lastly, the discrepancy between the hop distance, used in the Edmonds-Karp algorithm, and the shortest path distance, which is preferred, could lead to longer paths.
This discrepancy is lower for the HSPM method, due to its finer segmentation approach. 
More investigations are needed to make definitive statements about the effects of this discrepancy.
A combination of these shortcomings explains the apparent systematic underestimation of $Q$ by the ISPM method (Figure~\ref{fig:Q_comparison_seg}). 

%%%
%%% SUBSECTION - ACCURACY OF THE SENSITIVITY ANALYSIS
%%% 
\subsection{Accuracy of the sensitivity analysis of 1000 stochastic DFN}
%%% HSPM 
The high accuracy of the HSPM method found during the 100 stochastic DFN investigation (Figure~\ref{accuracy}) is reproduced in the 1000 case sensitivity analysis (Figure~\ref{CDF}), as it captures the predictions of the DS method for both the cumulative density function (CDF) and the corresponding histogram.

%%% ISPM
The ISPM method's CDF and associated histogram are shifted to the lower flow values, $Q$, and, although they cover the same flow-rate range, show higher probabilities for flow rate values between 0.2 and 0.4.
This shift to lower flow rates for the ISPM method results from its tendency to underestimate flow rates as observed in the 100 stochastic DFN investigation.

Naturally, the range in flow rates, computed for the different fracture densities and geometric constellations, is expected to increase if the investigation's parameter space is increased, as variable apertures across fractures would lead to more complicated flow paths, and non-percolating networks have a $Q$ value of zero. 

%%%
%%% SUBSECTION - COMPUTATIONAL COST
%%% 
\subsection{Computational cost} 
The computational cost for both methods is orders of magnitude lower compared to the DS method (Figure~\ref{Comp_cost_100_cases}).
This is apparent in the sensitivity analysis, where it took 7.59 days to obtain the DS results on a \unit[2.6]{GHz} XEON E5-2670 server processor, compared to 69 and 43 minutes, respectively, for the HSPM and ISPM methods on a \unit[2.9]{GHz} I7-3520M laptop processor.
An additional reduction of this computational cost is expected from further optimization of the code using specialized algorithms and optimized implementations on multi-thread or multi-core architectures \citep{myre2011performance,walsh2009accelerating}. 

The higher accuracy of the HSPM method, compared to the ISPM method, comes at the price of additional computational cost.
Because the HSPM method generates more edges and nodes, compared to the ISPM method (Figure~\ref{vert_edge}), HSPM calculations spend more time building the graph, performing calculations on the graph, and extracting results from the graph.
This difference in computational cost reduces with an increase  in the number of fractures, however, indicating that the current implementation of the HSPM method scales better than the current implementation of the ISPM method.

One area, where a straightforward reduction of the computational cost can be achieved for the developed methods is, when they are used for simulations, where the DFN geometry stays the same, but changes are made to the apertures, fluid properties, and/or the scale of the DFN.
When the geometry does not change, such investigations do not need to extract a new graph, but only need to change the edge values on an existing graph and extract the new result, thereby reducing a large part of the computational cost.

%%%%%%%%%%%%%%%%%%%%%%%%%%%%%%%%%%%%%%%%%%%%%%%%%%%%%%%%%%%%%%%%%
%%%
%%% CONCLUSION
%%%
\section{Conclusions} \label{sec:conclusion}
%%%%%%%%%%%%%%%%%%%%%%%%%%%%%%%%%%%%%%%%%%%%%%%%%%%%%%%%%%%%%%%%%

Two approaches for rapidly calculating the rate of fluid flow in discrete fracture networks (DFNs) have successfully been developed using existing combinatorial optimization algorithms: the Hanan Shortest Path Maxflow (HSPM) method and the Intersection Shortest Path Maxflow (ISPM) method. After comparing the results of simple test cases with direct simulations (DS), an investigation of 100 stochastic DFNs, with a range of fracture densities, was performed.
The flow rates predicted by the HSPM method, which represents parts of fractures (segments) as nodes, are on average \unit[5.03]{\%} lower than the results obtained from the DS.
The ISPM method, which represents fracture intersections as nodes, is less accurate (average difference of \unit[16.38]{\%}, compared to the DS).

The most important improvement that can be made to increase the accuracy of both methods arises from the behavior of the employed Edmonds-Karp algorithm \citep{ahuja1993network}.
Because of fundamental differences between fluid flow in fractures and ``combinatorial flow" on a graph, this algorithm can ignore parts of fractures or parts of fracture paths that may carry non-negligible fluid flow.

Both methods are orders of magnitude faster than DS, with the ISPM method being faster than the HSPM method.
The computational cost of the HSPM method, however, is expected to scale better with increases in the number of fractures.

A wide range of DFN properties can be investigated using these methods by only changing the edge weights on an existing graph, such as investigating different aperture distributions using the same DFN geometry.
This further reduces the computational cost as the graph only needs to be extracted once.
These methods additionally do not suffer from meshing issues, as traditional meshing is not required.

The low computational cost and relatively high level of accuracy make these methods excellent candidates for rapid screening, as shown by a sensitivity analysis of fracture density and fracture constellation, employing 1000 stochastic DFNs. The methods also facilitate
in-depth uncertainty quantifications of fluid flow through DFNs, thereby enabling the assessment of high- and low-probability events, such as very little or very large fluid flow for a given fluid pressure gradient. 
Such characterizations additionally enable assessments of DFN properties that have a strong influence on such high- and low-probability events.
\newpage
%%%%%%%%%%%%%%%%%%%%%%%%%%%%%%%%%%%%%%%%%%%%%%%%%%%%%%%%%%%%%%%%%
%%%
%%% NOMENCLATURE
%%%
\section{Nomenclature \label{nom}}
%%%%%%%%%%%%%%%%%%%%%%%%%%%%%%%%%%%%%%%%%%%%%%%%%%%%%%%%%%%%%%%%%
\begin{longtable}{p{.165\textwidth}p{.125\textwidth}l}
  \caption{List of symbols}
   \endfirsthead
  \\
\multicolumn{3}{c}%
{{\footnotesize \tablename\ \thetable{} (continued): List of symbols}} \medskip\\ 
\toprule
Symbol & Unit & Description\\
\midrule
  \endhead
  
  \bottomrule \multicolumn{3}{r}{{\footnotesize Continued on next page}}
\endfoot

\bottomrule
\endlastfoot

      \toprule
      Symbol & Unit & Description\\
      \midrule
      $a$ 				& $\unit{m}$ 				& 	Fracture aperture          				\\
      $\text{CPU}_{\text{DS/HSPM/ISPM}}$ & $\unit{s}$ 		& 	Computational cost of DS, HSPM, and ISPM calculations	\\
      $c_e$, $cr_e$ 	& - 						& 	Edge capacity and residual edge capacity\\
      $D$ 				& - 			            & 	Mesh-refinement coefficient				\\
      $dx, \,dy,\,dz$ 	& $\unit{m}$ 				& 	Spatial discretization         			\\ 
      $E$, $E_r$		& -							&  	Set of edges on a graph and on a residual graph \\
      $e$, $e'$, $\#e$ 		& - 						&	Edge index, reverse edge index and number of edges		\\
      $el$ 				& 							& 	Cubic element in a DS 					\\
      $F$				& -							& 	Face of a cubic element in a DS 		\\
      $f$, $f_e$		& -							& 	Flow on a graph and flow on an edge 	\\
      $G$, $G_r$ 		& - 						& 	Graph network and residual network	 	\\
      $g$ 				& $\unit{m/s^2}$	        &	Gravitational acceleration 				\\
      $K$ 				& $\unit{m^3/(Pa\cdot s)}$  &	Hydraulic conductance 					\\
	  $L_e$, $L_p$ 		& $\unit{m}$  				&	Edge length and path length				\\
      $L_{\text{short},p}$ & $\unit{m}$ 			& 	Shortest distance between a path's start and end points \\
      $l$ 				& $\unit{m}$            	&	Fracture length							\\
      $n$ 				& -							&	Upper limit for sums					\\
      $P$				& $\unit{Pa}$           	&	Fracture fluid pressure 				\\
      $p$				& - 						&  	Path index 								\\
      $Q$, $Qp$ 		& $\unit{m^3/s}$			&	Fluid volume rate for the network and for a path \\
      $s$				& - 						&	Source node								\\
      $T$ 				& $\unit{s}$ 	      		&	Time 									\\
      $t$				& - 						& 	Sink/target node 						\\
      $V$, $v$				& -							&  	Set of vertices on a graph and number of vertices 				\\
      $W$ 				& $\unit{m}$ 	    	 	&	Fracture width 							\\
      $w$, $w_e$ 		& $\unit{m}$ 	    		&	Width assigned to an edge  				\\
      $w_\text{harm}$ 	& $\unit{m}$		   		&	Harmonic mean of edge widths 			\\
      $\Delta z$ 		& $\unit{m}$ 				& 	Vertical distance against gravity		\\
      $\alpha$			& -							& 	Smallest residual capacity on an augmenting path \\
      $\beta$ 			& - 						& 	Smallest flow result on a path  		\\
      $\delta$ 			& $\unit{m}$    			&	Mesh size 								\\
      $\varepsilon_\text{HSPM}$, $\varepsilon_\text{ISPM}$ & \% 				& 	Discrepancy of DS and HSPM results, and DS and ISPM results\\ 
      $\mu$ 			& $\unit{Pa \cdot s}$    	&	Dynamic fluid viscosity					\\
      $\rho$ 			& $\unit{kg/m^3}$           &	Fluid density 							\\ 
      $\tau$ 			& -			           		&	Correction factor 						\\
\end{longtable}
%%%%%%%%%%%%%%%%%%%%%%%%%%%%%%%%%%%%%%%%%%%%%%%%%%%%%%%%%%%%%%%%%

%%%%%%%%%%%%%%%%%%%%%%%%%%%%%%%%%%%%%%%%%%%%%%%%%%%%%%%%%%%%%%%%%
%%%
%%% ABBREVIATIONS USED IN THE TEXT
%%%
\begin{table}[!ht]
	\centering
	\caption{Abbreviations used in the text}
	\label{abbreviations}
	\begin{center}
      	\begingroup
		\setlength{\tabcolsep}{20pt} % Default value: 6pt
		\begin{tabular}{ll}
			\toprule
            DFN  & Discrete fracture network                   \\
            DS   & Direct simulations                          \\
			HSPM & Hanan Shortest Path Maxflow approach        \\
			ISPM & Intersection Shortest Path Maxflow approach \\
     		\bottomrule                                
		\end{tabular}
        \endgroup
	\end{center}
\end{table}
%%%%%%%%%%%%%%%%%%%%%%%%%%%%%%%%%%%%%%%%%%%%%%%%%%%%%%%%%%%%%%%%%

%\newpage
%%%%%%%%%%%%%%%%%%%%%%%%%%%%%%%%%%%%%%%%%%%%%%%%%%%%%%%%%%%%%%%%%
%%%
%%% ACKNOWLEDGMENTS
%%%
\section{Acknowledgments}
%%%%%%%%%%%%%%%%%%%%%%%%%%%%%%%%%%%%%%%%%%%%%%%%%%%%%%%%%%%%%%%%%

Alex Hob\'e thanks the Master Scholarship Program of ETH Zurich for financial support.
Daniel Vogler gratefully acknowledges funding by the Swiss Competence Center for Energy Research - Supply of Electricity (SCCER-SoE).
Martin Saar thanks the Werner Siemens Foundation for their endowment of the Geothermal Energy and Geofluids group at ETH Zurich, Switzerland. 
Any opinions, findings, conclusions, and/or recommendations expressed in this material are those of the authors and do not necessarily reflect the views of ETH Zurich or the Werner Siemens Foundation.

%%%%%%%%%%%%%%%%%%%%%%%%%%%%%%%%%%%%%%%%%%%%%%%%%%%%%%%%%%%%%%%%%
%%%
%%% BIBLIOGRAPHY
%%%
%\section*{References}
\bibliographystyle{model5-names}%\biboptions{authoryear}
%\bibliography{bibliography/bibliography.bib}

%%%%%%%%%%%%%%%%%%%%%%%%%%%%%%%%%%%%%%%%%%%%%%%%%%%%%%%%%%%%%%%%%

\end{document}